# Open Source Energy System Modeling Using Break-Even Costs to Inform State-Level Policy: A North Carolina Case Study

Binghui Li *, Jeffrey Thomas, Anderson Rodrigo de Queiroz, Joseph F. DeCarolis

Department of Civil, Construction, & Environmental Engineering
North Carolina State University
Campus Box 7908, Raleigh, NC 27695-7908
United States

**KEY WORDS**

**ABSTRACT**

Rigorous model-based analysis can help inform state-level energy and climate policy. In this study, we utilize an open-source energy system optimization model and publicly available datasets to examine future electricity generation, $CO_2$ emissions, and $CO_2$ abatement costs for the North Carolina electric power sector through 2050. Model scenarios include uncertainty in future fuel prices, a hypothetical $CO_2$ cap, and an extended renewable portfolio standard. Across the modeled scenarios, solar photovoltaics represent the most cost-effective low-carbon technology, while trade-offs among carbon constrained scenarios largely involve natural gas and renewables. We also develop a new method to calculate break-even costs, which indicate the capital costs at which different technologies become cost-effective within the model. Significant variation in break-even costs are observed across different technologies and scenarios. We illustrate how break-even costs can be used to inform the development of an extended renewable portfolio standard in North Carolina. Utilizing the break-even costs to calibrate a tax credit for onshore wind, we find that the resultant wind deployment displaces other renewables, and thus has a negligible effect on $CO_2$ emissions. Such insights can provide crucial guidance to policymakers weighing different policy options. This study provides an analytical framework to conduct similar analyses in other states using an open source model and freely available datasets.

## INTRODUCTION

Many US states have proposed plans to address the climate change threat [1]. North Carolina is the ninth most populous state, the fourteenth largest $CO_2$ emitter (2014) [2] in the United States, and the first state in the Southeast to adopt a Renewable Energy and Energy Efficiency Portfolio Standard (REPS) [3]. The NC REPS requires investor-owned utilities to meet at least 12.5% of their electricity demand through renewable energy resources or energy efficiency measures by 2021 [3]. Three carve outs, representing minimum shares of specific fuel types, were also defined: 0.2% solar by 2018, 0.2% swine waste by 2020, and 900,000 MWh of poultry waste by 2016 [3]. The deployment of solar PV has far exceeded the solar carve out, with 4.3% of North Carolina's total generation supplied by solar [4]. This high level of solar PV deployment is largely due to the rapid decline in investment costs and favorable contract terms for third party, utility-scale solar under the Public Utilities Regulatory Policies Act (PURPA). However, electric power producers in the state have had difficulty meeting the swine and poultry waste targets, and have repeatedly filed joint petitions to the North Carolina Utilities Commission (NCUC) seeking relief and delay [5].

These outcomes illustrate the challenge that state policymakers face in developing policy that balances environmental performance, affordability, and stakeholder interests. Debates over energy-related policies and incentives within the state persist, often in the absence of sound, rigorous analysis available to the public. The same is true in many other states. Filling this analytical need to prospectively evaluate policy is now critical since comprehensive federal action to mitigate climate change is not imminent, and responsibility has fallen to states.

Energy system optimization models (ESOMs) represent a self-consistent framework for evaluation that can be used to probe the effects of potential policy while considering future uncertainty. ESOMs are already a crucial tool in long-term energy planning and policy making at regional to national scales, and in recent years numerous models have been developed and applied [6–15]. ESOMs can also help planners at the state level [16]. A key advantage of such models over simple calculations is their ability to capture dynamic technology interactions across the modeled system, which can have a significant effect on the generation mix over time. Capturing such interactions is also critical when assessing the relative cost-effectiveness of different technologies.

The break-even point, sometimes referred as grid parity point [17,18], indicates the point at which the delivered cost of electricity from a given technology reaches a target, often assumed to be the prevailing cost of grid electricity in a particular region. Break-even cost is a useful financial metric because it indicates how close specific technologies are to achieving cost parity and therefore deployment, relative to conventional technologies. The break-even cost is typically determined based on comparisons of specific financial metrics, such as net present value [19], internal rate of return [20], and levelized cost of electricity [21], which have been adopted by a wide range of studies [22–26]. However, an increasing number of studies call these metrics into question due to their oversimplified assumptions, lack of uniform standards, and failure to account for grid dynamics that can affect break-even cost [17,21,27–30]. These studies suggest break-even costs should be evaluated in a market-based framework that considers the dynamic interactions between technologies with different dispatch characteristics meeting time-varying demand. ESOMs are well-suited to identifying break-even points in a market context, but such analysis has not been a focus in previous work. While ESOMs often incorporate various forms of sensitivity analysis to determine how changes in input parameters affect outputs of interest [31–37], this is the first application of an ESOM to formally identify break-even capital costs.

In this study, we employ an ESOM called Tools for Energy Model Optimization and Analysis (Temoa) [38] to conduct state-level analysis of North Carolina's electric sector through 2050. A key innovation in this work is the technique used in calculating break-even costs, which are directly derived from ESOM solutions, and thus reflect the system-level values of each technology. Such information is particularly useful since state energy policy often aims to incentivize the deployment of technologies that are not currently cost-effective. Information on break-even investment costs that vary over time and under different scenarios can provide critical insight by helping policy makers to develop targets and financial incentives.

We begin by examining future electricity development pathways for North Carolina while considering fuel price uncertainty, a hypothetical $CO_2$ cap, and an extended REPS. Next, we examine the break-even costs under these different scenarios and use them to inform the consideration of a hypothetical tax credit under the extended REPS. Finally, we examine total electric sector $CO_2$ emissions and $CO_2$ abatement costs under all modeled scenarios. The objective of this analysis is twofold: develop policy-relevant insights that are both specific to North Carolina and generalizable to other states, and demonstrate an open source analytical framework that can be used to explore policy options in different states, regions, or countries. A key feature of Temoa is publicly archived source code and data, which enables third party replication and can serve as the basis for further analysis and exploration. The model source code and data are available through GitHub [39], and an exact copy of the files used to produce this analysis is archived through Zenodo [40].

## MODEL AND DATA

**Model Overview.** We use Temoa [38], an open-source, Python-based ESOM, to examine electric sector capacity expansion and associated emissions from 2015 to 2050. Temoa represents an energy system as a process-based network in which technologies are linked together by flows of energy commodities.

Each process is defined by an exogenously specified set of techno-economic attributes such as investment costs, operations and maintenance costs, conversion efficiencies, emission rates, and availability factors. Temoa is similar in structure to other ESOMs such as MARKAL [41], TIMES [9], MESSAGE [42], and OSeMOSYS [43].

Temoa is formulated as a linear program that minimizes the total system cost of energy supply over the user-specified time horizon, subject to both physical and operational constraints and user-defined constraints. Physical and operational constraints include conservation of energy at the individual process level, the global balance of commodity production and consumption, and the satisfaction of end-use demands. User-defined constraints include emission limits, maximum technology growth rates, and bounds on technology capacity and activity. Temoa minimizes the total system-wide cost of energy supply by optimizing the installation of new capacity and utilizing both new and existing capacity to meet demand. The complete algebraic formulation of Temoa is presented in Hunter et al. [38].

**Temporal Considerations.** In this study, the model time horizon spans 2015 to 2050, with each period consisting of five years. The results for each year within a given period are assumed identical. Note that although 2015 is a historical year, the first optimized period spans 2015 to 2019 and therefore the optimization results differ slightly from historical values.

NC electricity demand is projected to grow at 1.2% annually between 2015 and 2030 [44], based on forecasts from the Integrated Resource Plans (IRPs) of Duke Energy Progress and Duke Energy Carolinas [45], which constitute the largest utility serving North Carolina, as well as Dominion Energy, another electricity utility whose service territory includes the northeastern corner of North Carolina [46]. We extend this annual growth rate to 2050, and use the historical NC electricity consumption in 2015 as the base year value, as displayed in Table S2 and Figure S1.

ESOMs typically represent intra-annual variations in energy supply and end-use demands by dividing one year into a limited number of time slices that represent combinations of different seasons and times-of-day. Modeling supply and demand with fine-grained temporal resolution is necessary to capture the energy and capacity value of variable renewable energy sources [47,48]. Several papers attempt to model long time horizons with sufficient temporal detail to capture power sector operation in ESOMs [8,11,12,49]. In this study, one year is divided into 96 time slices: 4 seasons, with each season including 24 times-of-day to create a representative hourly profile for each season. This configuration allows us to capture average hourly variations in renewable resource availability and electricity demand. The load in each time slice comes from seasonal average load of that time-of-day, which is drawn from historical hourly electricity load in 2014 [50]. All scenarios utilize the same fixed, exogenously specified demand profile.

In addition, two constraints capture temporal aspects of power system operation: a system-wide reserve margin constraint and a ramp rate constraint. The reserve margin constraint requires that the total system capacity value must exceed the peak hourly demand by at least 15% [45] during each period to ensure adequate capacity reserve to meet demand during plant outages. Technology-specific capacities are multiplied by a capacity credit in the reserve margin constraint, where the capacity credit represents the fraction of capacity that can be relied on during peak demand periods. The assigned capacity credits are 5% for solar PV [51], 20% for onshore wind [51], 35% for offshore wind [52]; the remaining capacity credits for dispatchable generators are drawn from NERC [53]. Solar PV receives a low capacity credit due to limited solar availability during cold winter mornings when the system reaches its peak [51]. For simplicity, we assume that the capacity credit remains constant through time, though previous work indicates that the capacity credit of wind and solar declines with increasing penetration [54,55]. The ramp rate constraint requires that the change in electricity generation from a specific technology between two adjacent time slices must be bounded by its ramping capabilities. The mathematical formulation of these constraints is provided in the Supporting Information.

**Technology Cost and Performance Data.** The North Carolina electric sector is modeled as a single region and does not include a representation of the transmission network. Net interstate trade has constituted less than 10% of NC's total electricity supply [56] and is not included in this analysis. We performed an offline analysis of hourly imports and exports to the Duke Energy Progress system, which constitutes most of North Carolina, and did not observe large seasonal or diurnal variations in electricity trade that would have a significant effect on capacity expansion. We consider 28 electricity generating technologies, which can be categorized into nine groups based on their primary fuel types: natural gas, coal, diesel, uranium, biomass, geothermal, solar, wind, and hydro. Combustion technologies are defined based on their primary sources of power, including steam turbines, combustion turbines, or combined-cycle turbines. Advanced natural gas combined cycle and coal-fired steam with carbon capture and sequestration (CCS) plants are included, and state-of-art $SO_2$, $NO_X$ and $CO_2$ emissions control retrofits are also available for both existing and future units. Nuclear technologies include both conventional light water reactors (LWRs) and LWR-based small modular reactors (SMRs) as an advanced alternative. We consider three groups of renewable technologies: solar PV, wind, and biomass. Solar PV is further split into residential and utility-scale PV, differentiated by their investment costs and capacity factors. Wind power is categorized into onshore and offshore. Due to limited resource potential in North Carolina, onshore wind is capped at 5 GW in total [57]. Biomass-based integrated gasification combined-cycle (IGCC) is also included. Consistent with previous work [6,58], we model the input feedstock as a composite of corn stover, energy crops (grassy and woody), urban wood waste, agricultural, forest, and primary mill residues. Since end-use energy efficiency (EE) is a part of the current REPS, we model it as a generic technology. A literature review indicates EE costs ranging from 29 − 258 $/MWh [59–66], and the median variable cost of 43 $/MWh is used in this dataset. In addition, we consider four utility-scale electricity storage systems: lithium-ion battery, zinc-carbon battery, flow battery, and compressed air energy storage. The model treats the time slices as an ordered set and optimizes both the charge-discharge capacity and amount of energy stored or dispatched each time slice. In each model time period, the storage charge level is initialized to zero and must be fully discharged by the end of the period. The storage duration is fixed at 4 hours for simplicity.

Existing capacities are drawn from EIA Form 860 and calibrated with EIA's state electricity profile [56,67]. Technical parameters for most technologies, including costs, performance, and emission factors, are taken from EPA's MARKAL 2016 database supplemented by EIA's Annual Energy Outlook 2017 [6,68]. Capital costs of new electricity generating technologies are drawn from NREL's 2018 Annual Technology Baseline [69]. Solar PV costs in 2015 are taken from a market report from NREL [70]. A complete list of technologies and their techno-economic parameters are provided in Section 2 of the Supporting Information.

**Modeled Scenarios.** All modeled scenarios include EPA's Cross-State Air Pollution Rule (CSAPR), which limits $SO_2$ and $NO_X$ emissions [71] as well as the current REPS law. We model the current NC REPS with annual percentage targets for minimum renewable electricity generation and energy efficiency (EE). It requires at least 12.5% of electricity generation from renewable sources or EE in 2021 and beyond. We include the solar carveout, but not the ones for swine or poultry waste given their low targets. Consistent with the REPS law, the maximum allowable EE fraction starts at 25% in 2015, reaches 40% in 2025 [3], and remains fixed at 40% thereafter. The percentage requirements by model time period is detailed in Figure S10 of the Supporting Information.

Future electricity system pathways can be affected by several factors, but we choose to focus on three high-level issues: natural gas prices, an extended REPS, and a limit on $CO_2$ emissions. In North Carolina, as elsewhere, low natural gas prices enabled by hydraulic fracturing have led to a rapid transition away from coal and towards combined-cycle gas turbines (Figure S1 in the Supporting Information). Thus, the future generation mix will be sensitive to realized natural gas prices. In addition, there are active and ongoing discussions about future energy and climate policy, including Executive Order 80, which aims to reduce statewide $CO_2$ emissions by 40% below 2005 levels by 2025 [72].

In this study, we utilize future fossil fuel price projections from EIA's Annual Energy Outlook 2017 (AEO 2017) [68]. Fuel prices from three AEO2017 scenarios are selected to encompass the full range of natural gas prices included in AEO 2017: Low Oil and Gas Resource, Reference, and High Oil and Gas Resource, which have the highest, intermediate, and lowest natural gas prices, respectively. The price trajectories are shown in Figure S8 in the Supporting Information. As noted in the results, the large variation in projected natural gas prices has a significant effect on capacity deployment.

We also consider a hypothetical cap on $CO_2$ emissions. In addition to Executive Order 80, Duke Energy, the largest investor-owned utility in North Carolina, has committed to reducing its emissions 40% below 2005 levels by 2030 [73]. This goal is meant to be consistent with a scenario in which the world collectively limits climate change to no more than 2°C above pre-industrial levels [73]. In our analysis, we provide a linear extrapolation of this goal to achieve a 70% reduction below 2005 levels by 2050. We assume for simplicity that proportional reductions are undertaken by all states, and thus leakage effects across state lines are minimal. While Duke models their $CO_2$-constrained scenario as a carbon price, the assumed value is not publicly available in their integrated resource plan [45]. Historical $CO_2$ emissions from the electric sector and the modeled future emission limits from 2025 to 2050 are given in Figure S9 in the Supporting Information.

Finally, we consider a revised and extended REPS that represents a linear extrapolation of the current REPS target, reaching a 30% share of renewable electricity in 2050 (Figure S10 in the Supporting Information). The existing REPS has already been achieved, and active discussions about an extended REPS or clean energy standard are currently taking place. In the extended REPS scenario, the EE fraction of all renewable electricity is assumed to remain the same as in the current REPS, i.e., it starts at 25% in 2015 and reaches 40% in 2025. In this revised REPS, the target includes utility and residential solar PV, onshore and offshore wind, hydro, and biomass IGCC with no carve outs or additional financial incentives.

The scenario analysis therefore includes the following scenarios: no new policy with high (H), reference (R), and low (L) natural gas prices; the carbon cap with high (Cap-H), reference (Cap-R), and low (Cap-L) natural gas prices; and the extended REPS under reference level natural gas prices (REPS-R). We only include the extended REPS under reference natural gas prices, as results in the REPS and no cap scenarios under different fuel price projections are very similar. An additional REPS run is conducted with an investment tax credit (REPS-R-ITC) for wind, which is informed by the break-even analysis described below. Thus, a total of 8 scenarios are modeled. Table S12 in the SI summarizes the modeled scenarios.

**Break-even Analysis.** In most ESOMs, optimal solutions only consist of a subset of all available technologies, and they fail to reveal how close some technologies are to being deployed. We quantify the technology-specific break-even capital costs required to achieve deployment under the eight modeled scenarios. In this analysis, the break-even cost represents the capital cost at which a given technology will be deployed, all else equal. Examining break-even costs for technologies that are not deployed in a given scenario and period indicate the necessary capital cost reduction to make them cost-competitive. We focus on break-even capital costs because new low carbon technologies – including wind, solar, carbon capture and sequestration, and nuclear – are capital-intensive. In addition, capital costs also serve as a convenient metric, and can be readily compared with future cost projections from other sources [69,74]. Such information can help inform policy.

In this analysis, we use the reduced costs returned by Temoa's solver to estimate break-even costs. In linear programming, the "reduced cost" associated with a specific decision variable in the objective function is the amount by which its coefficient must improve before it can enter the optimal solution [75]. In Temoa, the reduced cost vector returned by the solver contains updated objective function coefficients associated with the technology-specific capacity variables that are not part of the initial optimal solution. These reduced cost coefficients indicate the level to which the fixed cost of each technology must drop – all else equal – to enter the solution. In simpler terms, the Temoa objective function represents a present cost calculation over the model time horizon. Our break-even cost calculation effectively estimates the required capital cost for each technology to make its present cost competitive with other technologies. This is not a static calculation, but rather determined endogenously by the model, as it depends on the dynamic interaction among all technologies meeting demand over time and subject to a set of performance constraints. A more detailed discussion on reduced cost and its relationship to break-even capital cost is provided in Section 4 of the SI. In addition, Section 4.3 in the SI compares results produced with our proposed method to a simple levelized cost of electricity comparison, and illustrates how the latter can lead to misleading insights by ignoring system-level constraints. Previous work points out the utility of reduced cost as a metric in energy system optimization models [76–79], but it has not been formally applied to quantify break-even costs.

One limitation of this approach is that break-even costs for a given technology and scenario will be contingent on all other cost and performance assumptions in the model. To address this limitation, we perform sensitivity analysis on the technology-specific investment costs. We perform this sensitivity in the L and Cap-H scenarios, which span the full range of break-even costs for each technology. Thus for a specific technology and scenario, the break-even costs are calculated three times, assuming the following for all generating technologies other than the one under consideration: (1) baseline capital costs, (2) a 20% increase in capital costs, and (3) a 20% decrease in capital costs. This uniform variation in capital cost across all generating technologies provides a simple way to roughly assess the relative sensitivity of technology-specific break-even costs to scenario assumptions and the capital costs of all other generating technologies.

## RESULTS AND DISCUSSION

**Capacity and Generation Mixes.** The electricity generation mix is shown in Figure 1. Low (L), reference (R), and high (H) fuel price scenarios affect the trade-off between natural gas, solar PV, and coal. In 2050, coal alone contributes over 50% of the total electricity generation in the H scenario, compared to less than 5% in the R and L scenarios. The H scenario results are consistent with previous studies [80,81], which report that in the absence of climate policy the US energy system continues fossil fuel use between 2010 and 2050. This observation aligns well with McCollum et al. [82], which found that the global energy system might see a future expansion of coal and low-carbon energy under high oil and natural gas prices. While the model results suggest that a limited resurgence of coal is possible under high natural gas prices, utilities are unlikely to make a 50-year investment in coal given the possibility of future climate policy.

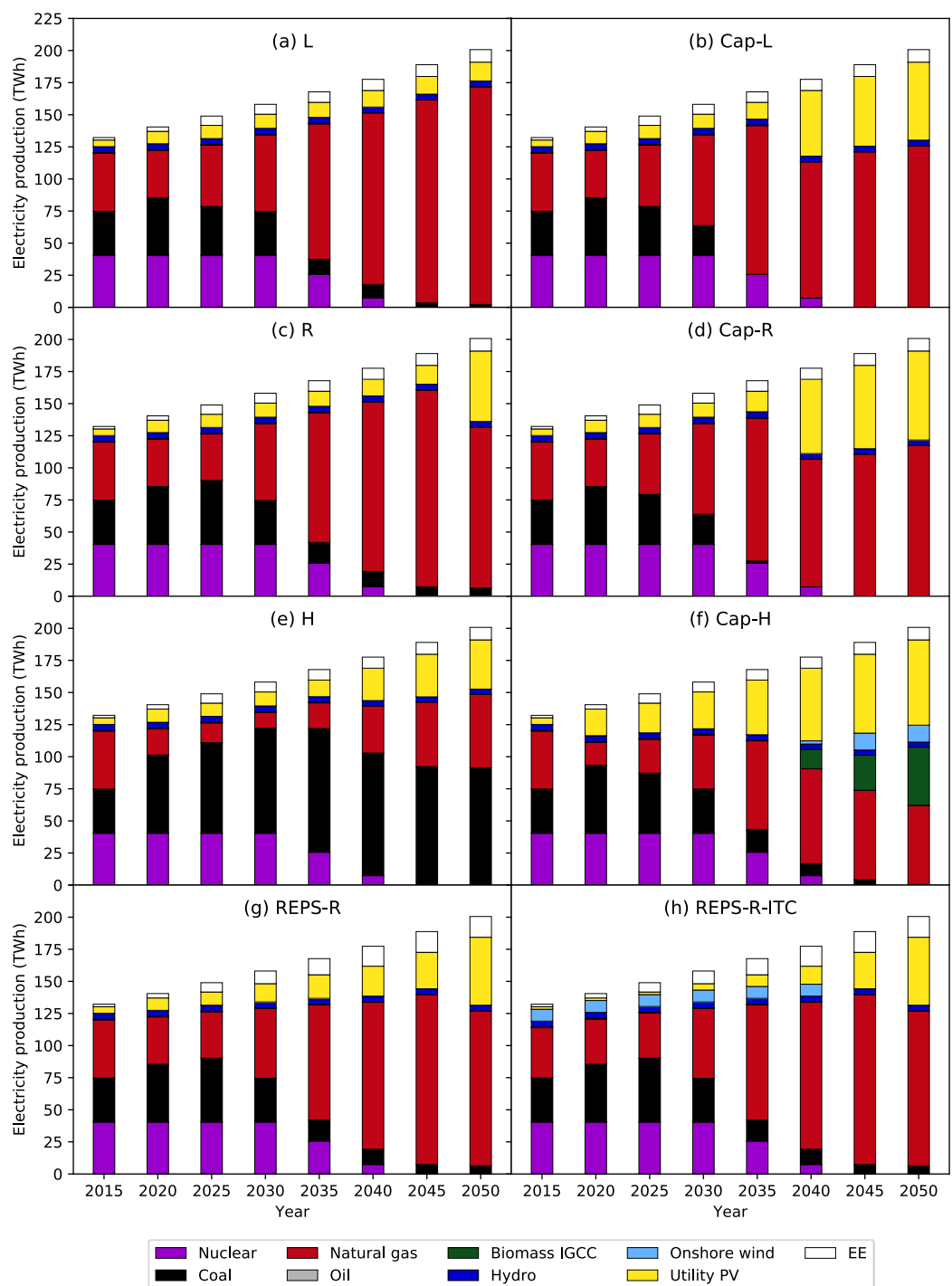

**Figure 1**. The NC electricity generation mix through 2050 under eight scenarios: (a) low natural gas prices [L], (b) carbon cap with low natural gas prices [Cap-L], (c) reference natural gas prices [R], (d) carbon cap with reference natural gas prices [Cap-R], (e) high natural gas prices [H], (f) carbon cap with high natural gas prices [Cap-H], (g) extended REPS with reference natural gas prices [REPS-R], and (h) extended REPS with reference natural gas prices and an investment tax credit for onshore wind [REPS-R-ITC]. Given the low cost of energy efficiency (EE) measures, it is used to the maximum extent under both the existing and extended REPS.

A direct trade-off between natural gas and renewables can be observed in the Cap scenarios. Natural gas contributes 31% of the 2050 electricity generation in the Cap-H scenario, whereas natural gas alone accounts for over 62% of electricity generation in the Cap-L scenario. In addition, the high natural gas prices in the Cap-H scenario produce a significant shift towards renewable energy. Solar PV, wind, and biomass collectively account for over 62% of the 2050 electricity generation in the Cap-H scenario, which represents the highest renewable penetration across all scenarios. The high natural gas prices in the Cap-H scenario lead to continued coal utilization, but coal is reduced to less than 1% of the generation mix by 2050. Although all scenarios have the same annual demands, the Cap scenarios typically have higher total capacities (Figure S11 in Supporting Information). The Cap scenarios include higher penetrations of wind and solar, which have relatively low capacity factors, and therefore require higher capacities to produce the equivalent amount of electricity as a conventional plant. Consistent with the Duke IRP [45], we consider CCS associated with pulverized coal and integrated gasification combined cycle (IGCC), but do not see its deployment in the carbon constrained scenario.

Figure 2 indicates that the future contribution of solar PV under all scenarios is significant, and the model results replicate the "duck curve" effect observed in California [83]. As shown in Figure 2f, the Cap-H scenario includes significant amounts of biomass, which are deployed to address the steeper ramps of net load resulting from a high solar PV penetration. The highest solar PV penetration in the Cap-H scenario creates steep ramps of net load during late-afternoon, especially in summer. As shown in Figure 2, net load in the Cap-H scenario in summer 2050 rises from 3 GW at 1pm to 25 GW at 8pm, creating a much steeper ramp than in the H scenario. However, natural gas only provides around 15 GW of ramping capacity due to high natural gas prices and therefore limited utilization of gas turbines. The model instead utilizes biomass IGCC, which is assumed to be a carbon-neutral process that enables fast-ramping through syngas combustion. While biomass IGCC alone provides over 7 GW of ramping capacity in 2050, NC biomass energy potential is estimated to be around 2 GW [57,84], which implies that an additional 5 GW of biomass resource must be imported from other states. Interestingly, storage is not deployed. We speculate that this may have to do with the fixed storage duration and the use of average 24-hour diurnal profiles per season that do not capture the full benefit of shifting supply from low to high demand periods. To better characterize the potential role of storage, additional work with higher temporal resolution of supply and demand and different parameterizations of storage technology is required [85]. For example, recent work using a model with hourly resolution indicated cost-effective deployment of lithium-ion battery storage in North Carolina by 2030 [86].

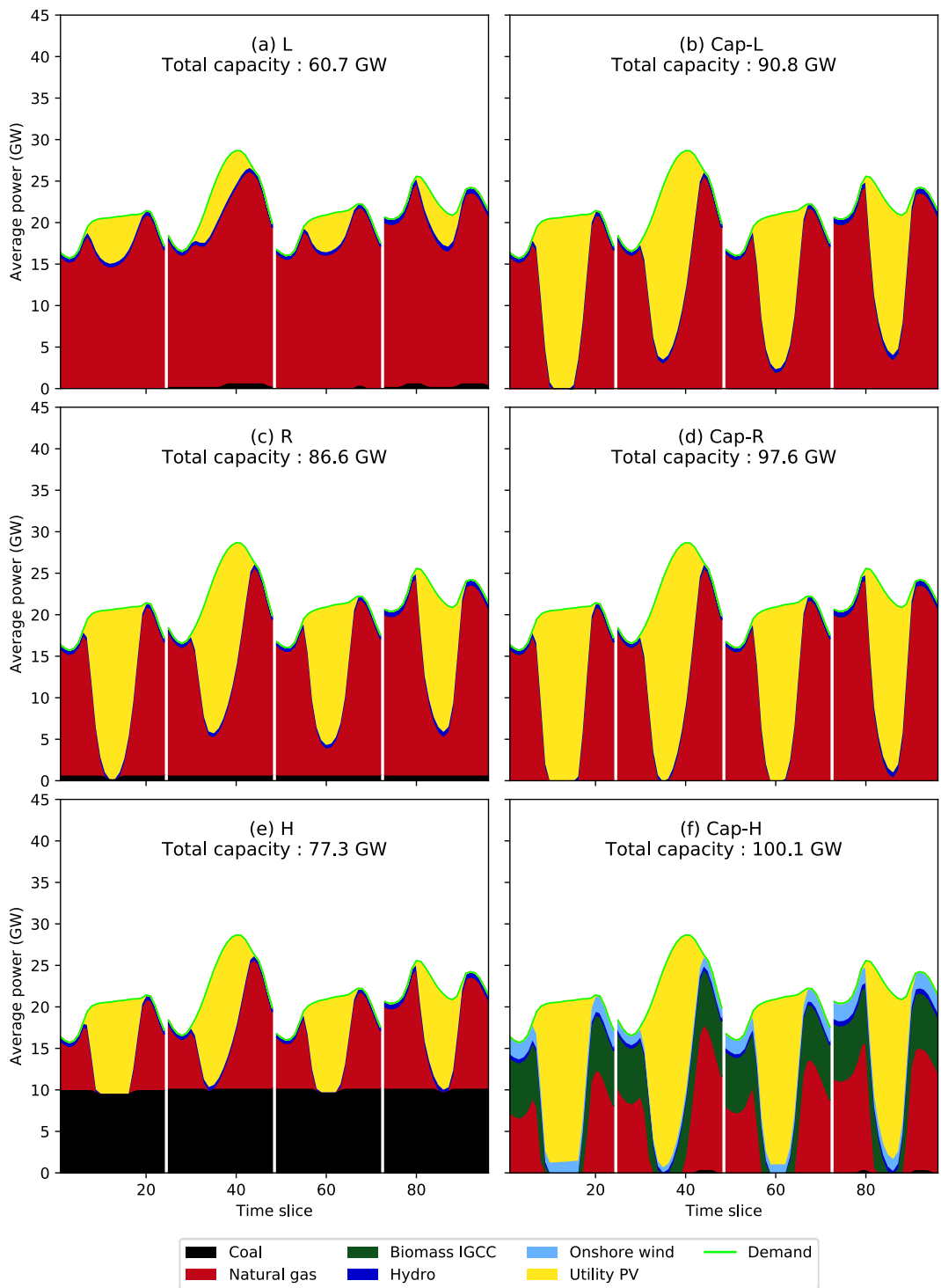

**Figure 2.** Average power production in 2050 from different sources in each time slice. Results from six scenarios are shown here: (a) L, (b) Cap-L, (c) R, (d) Cap-R, (e) H, (f) Cap-H. The seasons are ordered left-to-right beginning with spring. White gaps represent the boundaries between seasons, and each season is divided into 24 hour time slices. Note that net load, the total electric load minus wind and solar generation, is represented by the upper edge of the dark blue band.

**Break-even costs.** The trade-offs among low carbon options in the modeled scenarios largely revolve around utility-scale solar photovoltaics, biomass IGCC, and onshore wind. Decision makers may wonder how close to cost-effective other low carbon technologies may be under these different scenarios, and break-even costs can provide insight that helps inform policy. The break-even capital costs for six selected technologies are presented in Figure 3. In Figure 3, the upper edge of the light gray area represents the exogenously specified capital cost for each technology within the model. In the case of wind and solar technologies, the increase in investment cost from 2015 to 2020 is due the expiration of federal tax credits (see Section 3 in the SI for details). Declines in the specified capital costs over time represent the effect of technological learning, which are exogenous to the model. The markers occurring within the lighter grey area indicate that the investment costs must be reduced to that amount – all else equal – before the technology will be deployed during that period and scenario. When the markers overlap the upper edge of the lighter grey area, it implies that the technology has been deployed during that period. Dark gray and green bands around the L and Cap-H scenarios represent the variability in break-even capital cost for a given technology when all other technology capital costs are increased and decreased by 20%. Caution must be exercised when interpreting these results. The scenarios are selected to form a cost envelope, enabling a wide range of break-even costs across scenarios. However, the break-even costs plotted in Figure 3 are illustrative, and should be interpreted as discrete results drawn from a continuous space.

We chose to focus the break-even cost analysis on low carbon options, including renewables (onshore and offshore wind, residential solar PV, biomass IGCC) and nuclear (conventional light water reactors, small modular reactors). Utility-scale PV is excluded since it is already cost-effective in all modeled scenarios. Figure 3 indicates how the break-even costs vary under different scenario assumptions. In most cases, the break-even costs reach a maximum under the Cap-H scenario, which offers the best economic conditions under which to deploy low carbon technologies. Likewise, break-even costs are at a minimum under the L scenario, where low natural gas prices and no emissions limit make renewables and nuclear less cost-effective. The dark gray and green bands representing capital cost variation produce changes in the break-even cost that are comparable to the shift from one scenario to another. Variations in band thickness are also discernable. For example, the green band in Figure 3a is narrower than the grey band, indicating higher sensitivity of onshore wind breakeven costs to other technologies under the L scenario than under the Cap-H scenario. By contrast, the width of the green bands are wider than the grey bands in Figure 3c-d, indicating that breakeven costs of nuclear technologies are more sensitive to variations in the capital costs of other technologies if the $CO_2$ cap is in effect.

Figure 3c indicates that new light water reactors are cost-effective under the Cap-H scenario with a capital cost of 4500 \$/kW, but the break-even capital cost drops below 2000 \$/kW in the L scenario. Thus the cost-effectiveness of conventional nuclear is highly sensitive to both fuel prices and the presence of the $CO_2$ cap. Small modular reactors share a very similar break-even cost pattern to conventional light water reactors given their similar cost and performance characteristics. Because offshore wind is much more expensive than onshore wind, onshore wind is deployed in the Cap-H scenario and the capital cost reductions required in the other scenarios are smaller compared with offshore wind. However, given offshore wind's higher capacity factors, it can be cost-effectively deployed at a higher capital cost than onshore wind. Surprisingly, biomass IGCC exhibits negative break-even costs in the fuel price scenarios without a carbon cap, indicating that there is no break-even capital cost that makes it cost-effective in those scenarios. Even with zero capital cost for biomass IGCC, it still has high fixed operations and maintenance (O&M) costs: the present cost of the fixed O&M represents over 50% of its capital cost, but is less than 30% for most other technologies.

                                                                 

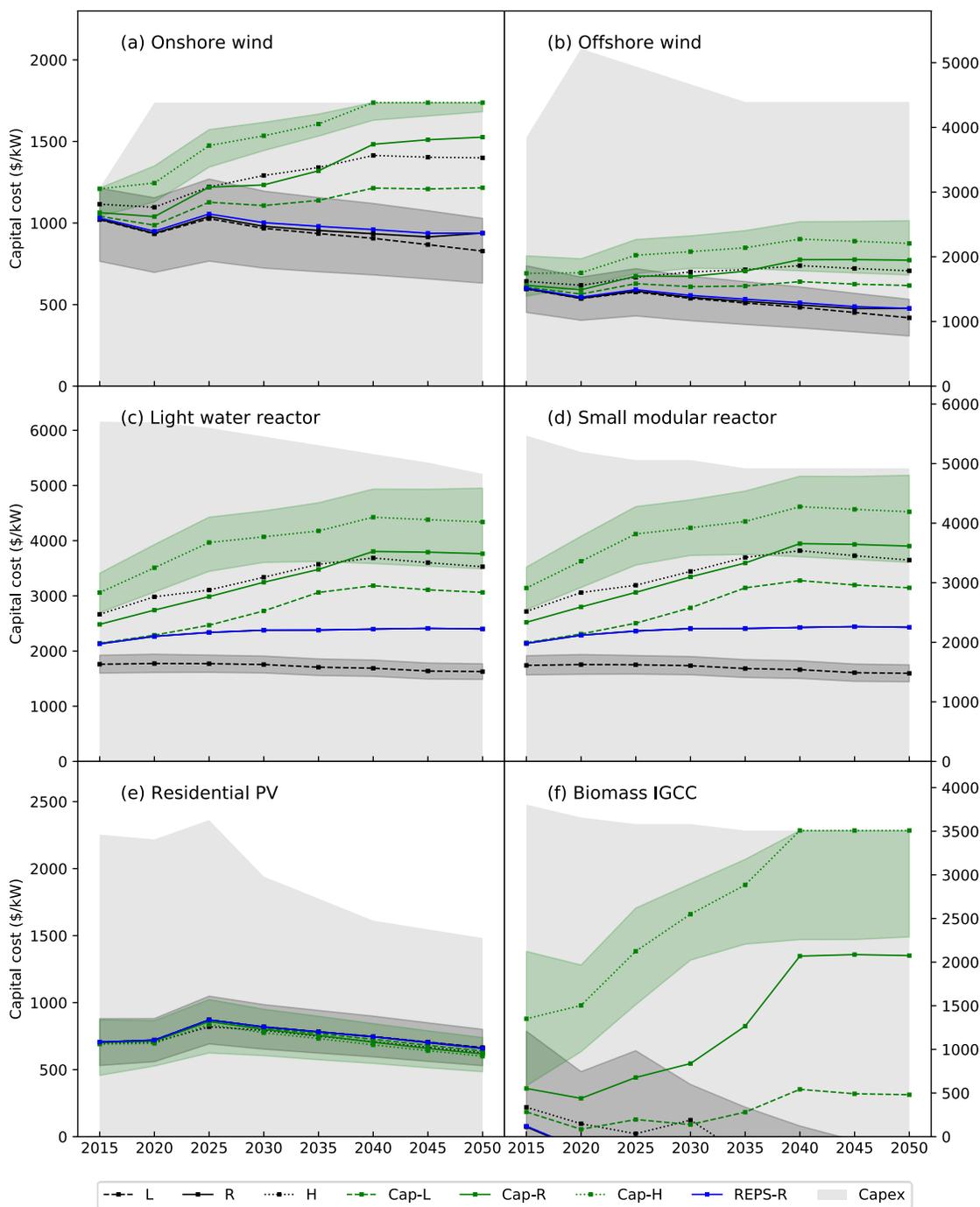

**Figure 3.** Break-even investment costs associated with (a) onshore wind, (b) offshore wind, (c) light water reactors, (d) small modular reactors, (e) residential solar PV, and (f) biomass integrated gasification combined-cycle. The dark gray and green bands represent the variation in break-even cost for a given technology when capital costs of all other generation technologies are simultaneously increased and then decreased by 20%. Note that in (c) and (d), the solid blue curve (REPS-R) is overlapping the solid black curve (R). In (f), the break-even costs of the L, R, H, and REPS-R scenarios are negative and are not displayed.

**The extended REPS scenario.** Suppose that state legislators decide to pursue the extended REPS (REPS-R). Compared with the R scenario (Figure 1c), the REPS-R (Figure 1g) includes additional solar PV generation, which displaces natural gas generation. Legislators might wish to consider an additional provision under REPS-R to promote diversity in renewables supply. Using the break-even cost information in Figure 3, further suppose that lawmakers decide to target the deployment of onshore wind. The break-even cost for onshore wind under the REPS-R scenario ranges from 930–1060 $/kW from 2020–2050 (Figure 3a), corresponding to a capital cost reduction between 32–39%.

This break-even cost information can be used in at least four different ways to inform the extended REPS. First, the required capital cost reduction can be used to calibrate an investment tax credit for wind to stimulate its deployment. Second, policy makers could decide to create a carve out for onshore wind, and the breakeven price could be used to set a cap on the incremental cost of wind Renewable Energy Certificates (RECs) that are allowed to be passed on to rate payers. Third, a carve out for wind can be created, and the break-even price used to calibrate an alternative compliance payment (ACP), which provides a price ceiling in the RECs market[87]. ACPs are penalty payments that a utility pays to the utilities commission or other governmental body if renewable energy goals are not met. As they are typically not recoverable from ratepayers, it is important to set the ACP at an appropriate level to encourage compliance with REPS laws. Wiser et al.[88] provide additional details about ACPs and values previously used in different U.S. states. Fourth, the break-even prices could be used to guide electricity system regulators in long-term energy auctions for renewables. While not applicable in the vertically integrated NC electric sector, these prices could potentially serve as the basis for designing appropriate subsidies to promote competition and achieve the lowest bids by generation source in competitive power markets[89].

For simplicity, we consider an investment tax credit for onshore wind – Option 1 above – by performing an additional run with an assumed 40% reduction in the wind capital cost, which is slightly higher than the 32–39% required reduction derived from Figure 3a. As indicated in Figure 1h, the proposed ITC allows onshore wind to achieve 6% of the 2040 generation mix under the extended REPS.

**$CO_2$ emissions and abatement costs**. The $CO_2$ emissions across all scenarios are shown in Figure 4. The $CO_2$ cap is binding across the three carbon constrained scenarios, and is represented by the green line in Figure 4. Only the H scenario shows a substantial rise in $CO_2$ emissions – 80% above the 2015 level in 2050 – driven by the resurgence of coal. In the L, R, and REPS-R scenarios, the emissions reduction between 2020 and 2035 reverses in later time periods due to demand growth coupled with higher natural gas utilization. Cumulative $CO_2$ emissions in the REPS-R scenario are 5% lower than the R scenario, indicating that the extended REPS only has marginal effects on $CO_2$ emissions. In addition, Figure 4 shows the $CO_2$ emission profile of the REPS-R scenario with and without the investment tax credit for wind are nearly the same, implying that the wind ITC leads to the deployment of wind at the expense of solar rather than increasing their overall combined deployment. Thus the wind ITC has negligible effects on fossil-based generation.

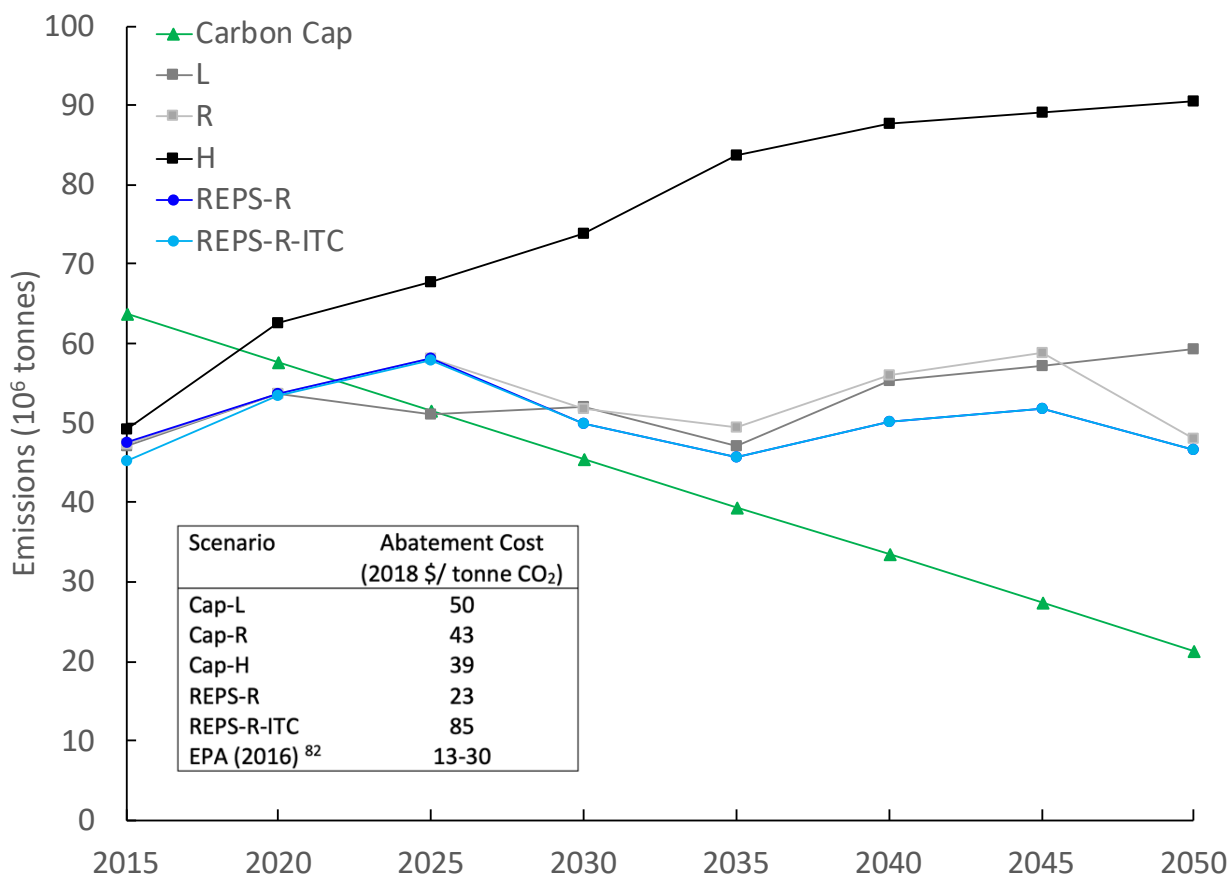

**Figure 4.** CO$_2$ emissions across all tested scenarios. The CO$_2$ cap is binding regardless of the prevailing fuel price projections in the cap scenarios. The table near the bottom left provides the average abatement costs calculated from the model results; the EPA social cost of carbon range from 2015-2050 at a 5% discount rate is included for comparison. All abatement costs are in 2018 dollars.

In addition, we calculate the average CO$_2$ abatement costs ($ per tonne CO$_2$) by computing the ratio of the difference in total costs to the difference in total emissions between each pair of scenarios consisting of a cap scenario and a scenario of the same fuel price without the cap. Similarly, the REPS-R scenario is compared with the R scenario. Therefore, the CO$_2$ abatement costs are proportional to increases in total costs, and inversely proportional to CO$_2$ emission reductions. The CO$_2$ abatement costs are included in Figure 4. The Cap-H scenario has the lowest average CO$_2$ abatement cost due to the proportionally larger CO$_2$ emission reduction than in the other two cap

scenarios. Although the $CO_2$ emission reductions in the Cap-L and Cap-R scenarios are similar, the cost increase from the L to the Cap-L scenario is higher than from the R to the Cap-R scenario due to the higher cost of replacing natural gas technologies with renewables in the L scenario. In the REPS-R-ITC scenario, we include the cost of the ITC in the calculation of abatement cost. Note that the ITC pushes the abatement cost above 80 \$/tonne $CO_2$ because it does not increase the level of emissions reductions. In such a scenario, policy makers must weigh whether such additional cost is worthwhile in light of other objectives, such as supply diversification and rural economic development. For comparison, Figure 4 also includes EPA's social cost of $CO_2$ (SC-$CO_2$) values from 2015 to 2050 at a discount rate of 5% [90], which is the same global discount rate used in our model. Though the modeling approach is different, the $CO_2$ abatement costs in the three cap and REPS-R scenarios are generally higher but overlap the EPA estimates.

**Policy Implications.** In North Carolina, solar PV is the most cost-effective low carbon technology, and would likely be used to meet future REPS or $CO_2$ cap requirements, consistent with past development. However, uncertainty in future capital costs could lead to different outcomes. The break-even cost represents the capital cost required to achieve the deployment of a technology that is not cost-effective under baseline assumptions. Our analysis indicates that the technology-specific break-even costs can vary significantly across scenarios. For example, the break-even nuclear costs vary by more than 20% of their baseline capital cost across some modeled scenarios. In addition, as indicated in Figure 3, variations in the capital cost of other generating technologies can shift the break-even cost of a given technology by roughly the same magnitude as switching between scenarios.

The break-even cost analysis presented here could be particularly helpful for states trying to formulate new energy or climate policy by allowing decision makers to compare the relative cost-effectiveness of different technologies under a wide variety of scenarios. Such information can be used in several ways. First, comparing break-even costs can help policy makers incentivize the deployment of technologies that deliver the highest public benefit at the lowest cost. Second, break-even costs can be used to determine alternative compliance costs, or the cap on a utility's allowable recovery of incremental REC costs from ratepayers. Third, break-even costs could be used to inform technology research and development aimed at achieving a particular cost target. Fourth, while the break-even costs in this analysis are based on future capacity expansion, it could be adapted to estimate the zero emissions credit levels necessary to maintain the existing fleet of nuclear over the next few decades. Several states, including Illinois, New Jersey, and New York have implemented zero emissions credits [91,92] for existing nuclear generators. The credits represent the zero emissions attribute of each megawatt-hour produced by a qualified nuclear power plant. Previous analysis indicates that the preservation of existing nuclear is a cost-effective carbon avoidance strategy [93].

Federal action on climate is not imminent. The United States plans to withdraw from the Paris Agreement, and the EPA Clean Power Plan has been repealed. However, several states have pledged to uphold the Paris Agreement [94], and many states have already taken actions, such as the Regional Greenhouse Gas Initiative [95] and California's cap-and-trade system [96]. The requisite policy planning would benefit from ESOM-based analysis, which can help states achieve their desired emissions targets cost-effectively. This analysis provides a blueprint that can be replicated in other states with a publicly available model.

**Caveats.** In this analysis, we try to place focus on our modeling approach and generalized insights that are robust to the model limitations and high future uncertainties. Nonetheless, several caveats are worth noting. First, scenario-specific results pertaining to future electricity generation, break-even costs, and emissions are dependent on the baseline assumptions used in the model. The capital cost sensitivity performed in break-even cost analysis demonstrates how technology-specific cost-effectiveness can change when the baseline capital cost of other technologies are varied. Though not presented here, we also tested a $\pm$ 40% range on other capital costs and observed a linear increase in the width of the gray and green uncertainty bands presented in Figure 3. Second, using a single solver occasionally returned anomalous results. Section 4 of the Supporting Information explains in detail the procedure we implemented to work around this difficulty. Third, while the increased number of time slices, capacity reserve constraint, and ramp rate constraint help constrain system performance within reasonable bounds, the model does not perform hourly unit commitment and dispatch [97] or consider the need for operating reserves [98]. Additional effort is required to increase the temporal resolution in ESOMs to better represent power system operation. Fourth, we assume a fixed exogenous demand across all scenarios. Price-responsive demands would lower electricity demand in cap scenarios with higher electricity prices, but would not fundamentally change the relative grid mix or technology cost-effectiveness. Finally, while we demonstrate the utility of break-even cost analysis, it represents one of many sensitivity and uncertainty analysis methods that should be brought to bear on model-based analysis aimed at informing policy [99].

## ASSOCIATED CONTENT

**Supporting Information** Additional constraints, including the reserve margin constraints and the ramp rate constraints; Overview to the electric system of North Carolina, the electricity consumption history and future demand predictions (Table S2 and Figure S1); Environmental regulations, emission limits (Figure S5); Fuel costs (Figure S8); Supplementary results, capacity mixes (Figure S11).

## AUTHOR INFORMATION

### Corresponding Author

*E-mail: bli6@ncsu.edu; phone: +1-919-348-5610; address: 2501 Stinson Drive, Raleigh, NC 27695-7908

### Notes

The authors declare no competing financial interest.

Approaches to Uncertainty Assessment in Energy System Optimization Models. *Energy Strateg. Rev.* **2018**, *21*, 204–217.

**Supporting Information**

Open Source Energy System Modeling Using Break-Even Costs to Inform State-Level Policy: A North Carolina Case Study

Binghui Li *, Jeffrey Thomas, Anderson Rodrigo de Queiroz, Joseph F. DeCarolis
Department of Civil, Construction, & Environmental Engineering, North Carolina State University Campus Box 7908, Raleigh, NC 27695-7908 United States

* Corresponding author: bli6@ncsu.edu

Number of pages: 46

Number of tables: 19

Number of figures: 11

# Table of Contents

## List of Tables

## List of Figures

# 1. Additional Constraints

The Temoa formulation includes several constraint sets, which are discussed in detail in Hunter et al.[1]. In this analysis, two additional constraint sets are added to account for short-term power system operations: the system-wide reserve margin constraint and the ramp rate constraints.

## 1.1. The Reserve Margin Constraint

The reserve margin constraint requires Temoa to build enough capacity in each period to satisfy the capacity reserve margin. The capacity reserve margin is defined by the North American Electric Reliability Cooperation (NERC) as the amount of unused electric capacity at the time of peak load [2]. The capacity reserve requirement ensures that utilities can meet peak demand at all times, thereby maintaining system reliability. Each balancing authority under NERC sets a reference level, and assesses the ability of utilities to meet this reliability metric. North Carolina falls under the Southeastern Reliability Cooperation (SERC) balancing authority, which sets a 15% reference reserve margin [3]. Utilities in North Carolina are required by statute [4] to submit integrated resource plans (IRPs), which outline their ability to meet future demand with new and existing resources [5]. Both Duke Energy Progress [6] and Duke Energy Carolinas [7] are required to meet the 15% minimum summer reserve margin target [2].

The constraint is defined in Equation 1. During each period $p$, the sum of the available capacity ($\mathbf{CAPAVL}_{p,t}$) of all reserve technologies, which are defined in the set $\mathbf{T}^{res}$, should exceed the peak load by $RES_c$, the reserve margin. Note that the available capacity is first discounted by the capacity credit ($cc_t$) and then summed up.

$$\sum_{t \in \mathbf{T}^{res}} cc_t \cdot \mathbf{CAPAVL}_{p,t} \cdot SEG_{s*,d*} \cdot C2A_t \geq DEM_{p,c} \cdot DSD_{s*,d*,c} \cdot (1 + RES_c),$$
$$\forall p \in \mathbf{P}^O, c \in \mathbf{C}^{res} \tag{1}$$

where, $(s*, d*)$ indicates the peak-load time slice, $c \in \mathbf{C}^{res}$ indicates the demand commodity that is subject to the reserve margin constraint, $C2A_t$ is the technology-specific conversion factor between capacity and activity, and $SEG_{s*,d*}$ and $DSD_{s*,d*,c}$ represent the fraction of time and the fraction of load within the peak-load time slice, respectively. It is important to note that the SERC-specified reserve margin of 15% must be adjusted to account for the demand representation in the NC dataset. Due to computational constraints, we represent 8760 hourly demand slices as 96 representative time slices in Temoa, which creates a discrepancy between the actual hourly peak load and the average peak load in the model (See Figure S2). As a result, we apply a 60% reserve margin in Temoa, which accounts for the difference between the real-world and modeled peak loads as well as the 15% reserve margin in SERC.

## 1.2. The Ramp Rate Constraint

The ramp rate constraint is utilized to limit the rate of electricity generation increase and decrease between two adjacent time slices in order to account for physical limits associated with thermal power plants. The formulation is given in Equation (2).

$$-r_t \cdot \mathbf{CAP}_{p,t} \leq \frac{\mathbf{ACT}_{p,s,d+1,t,v}}{SEG_{s,d+1} \cdot C2A_t} - \frac{\mathbf{ACT}_{p,s,d,t,v}}{SEG_{s,d} \cdot C2A_t} \leq r_t \cdot \mathbf{CAP}_{p,t}$$
$$\forall p \in \mathbf{P^O}, s \in \mathbf{S}, d, d+1 \in \mathbf{D}, t \in \mathbf{T^{ramp}}, v \in \mathbf{V}$$
(2)

In the above equation, $r_t$ represents the ramp rate limits for those technologies defined as ramping technologies ($t \in \mathbf{T^{ramp}}$); $\mathbf{CAP}_{t,v}$ and $\mathbf{ACT}_{p,s,d,t,v}$ represent the installed capacity and activity of technology $t$ deployed in vintage $v$, respectively. We assume for simplicity that technology ramp rates do not vary with technology vintage. Coal-fired and nuclear power plants are mostly regarded as baseload sources due to technical and economic limitations, while natural gas-fired units are usually employed as load-following units given their greater operating flexibility and faster response. The ramp rates used in this study, as shown in Table S1, are drawn from FERC [8], which is based on real-world samples in the PJM interconnection. In this study, only 60% of the existing coal capacity is subject to the ramp rate constraints, assuming they can be utilized as load-following units, while 40% is considered baseload and does not vary output through the day. This approach ensures that the coal units exceed their minimum run levels, which usually fall between 20% and 40% of their nameplate capacities [8,9]. Table S1 shows that natural gas units can ramp up or down by more than 100% within an hour.

Table S1. The ramp rate limits given by a sampling-based study for the generation fleet in PJM. Data drawn from Krall et al. [8]

| Tech name | Ramp up/down (%/min) | Ramp up/down (%/hr) |
|---|---|---|
| Coal | 0.7% | 42% |
| NG, combined | 1.8% | 108% |
| NG, combustion | 3% | 180% |

Although the cold start-up time of a natural gas unit can be significantly longer than one hour, the issues on start-up are not accounted for in this study. Therefore, we assume the ramp rate constraint only applies to coal-fired units, and all nuclear units are viewed as baseload. Other thermal power plants with low ramping capability, such as biomass-fired steam turbines, are excluded due to their insignificant share.

## 2. An overview of the electric power system in North Carolina

In 2015, total electricity generation in North Carolina was 128,000 GWh [10], as shown in Table S2. Electricity generation by source is provided in Figure S1 and Table S3. The North Carolina Energy Policy Council projects that the demand for NC electricity will grow at an annual rate of 1.2% between 2015 and 2030 [11], which equates to a 20% increase by 2030. Note that this growth rate is based on the IRPs of Duke Energy Progress [6] and Duke Energy Carolinas [6], collectively representing the largest utility serving North Carolina, which provides over 70% of total electricity sales [12]. Dominion Energy, another electricity utility that primarily serves the northeastern corner of North Carolina, projects a slightly higher growth rate for its NC territory. Dominion's total electricity sales to North Carolina were 4,428 GWh in 2015 [12], accounting for less than 4% of North Carolina's total electricity generation. Therefore, we assume that total NC electricity demand will increase at a yearly growth rate of 1.2% from 2015 to 2050, and the historical NC electricity consumption in 2015 is used as the base year. Table S2 shows the annual electricity demand projections for North Carolina, and Figure S1 shows both the historical NC electricity consumption by fuel source and the projection of total electricity demand.

Table S2. Annual electricity demand projections.

| Period | 2015 | 2020 | 2025 | 2030 | 2035 | 2040 | 2045 | 2050 |
|--------|------|------|------|------|------|------|------|------|
| Demand (TWh) | 128 | 136 | 145 | 154 | 163 | 173 | 184 | 195 |

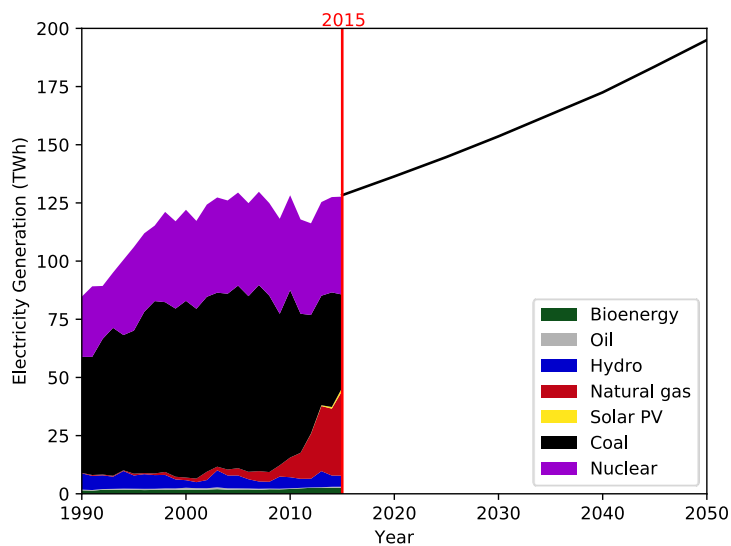

Figure S1. North Carolina's historical energy mix from 1990 to 2015 and electricity demand projections. The trajectory beyond 2015 represents projected electricity demand.

Table S3. North Carolina capacity and electricity generation mix in 2015 [a].

|  | Capacity (MW) | Generation (MWh) | Cap. % | Gen. % | Capacity Factor |
|--|---------------|------------------|--------|--------|-----------------|
| Bioenergy [b] | 487 | 2,589,372 | 1.6% | 2.0% | 61% |
| Coal [c] | 10,803 | 39,922,168 | 34.5% | 31.1% | 42% |
| Oil [d] | 403 | 434,587 | 1.3% | 0.3% | 12% |
| Hydro | 2,004 | 4,742,004 | 6.4% | 3.7% | 27% |
| NG [e] | 10,816 | 36,544,596 | 34.5% | 28.5% | 39% |
| Solar PV | 1,437 | 1,373,579 | 4.6% | 1.1% | 11% |
| Nuclear | 5,114 | 42,096,761 | 16.3% | 32.8% | 94% |
| Pumped Hydro | 86 | 94 | 0.3% | 0.0% | 0% |
| Other | 161 | 685,283 | 0.5% | 0.5% | 49% |
| Total | 31,311 | 128,388,444 | 100% | 100% | |

a. Data is from EIA's Form 860 [13] and calibrated with EIA's State Electricity Profile 2015 [14].
b. Including wood/biomass steam turbines, municipal solid waste (MSW) steam turbines, landfill gas (LFG) internal combustion engines and LFG gas turbines.
c. All coal used in North Carolina is bituminous.
d. Primarily diesel oil combustion turbines.
e. Includes both combined cycle (NGCC) and single cycle (NGSC).
   A critical challenge in the integration of renewable energy such as wind and solar PV is the

 

intermittency associated with their output. To represent variations in electricity output, one year is divided into multiple time slices. The time slice is assumed to be an indivisible unit within the model, over which the electricity generation from all technologies are assumed to be constant. In addition, electricity demand varies between time slices, and must be satisfied through electricity generation. Two parameters are employed to describe the time slices: the segment fraction ($SegFrac_{s,d}$) and the demand specific distributions ($DSD_{s,d,c}$). The segment fraction is the fraction of one year represented by each time slice, and the demand-specific distribution represents the fraction of annual demand that falls within each time slice.

In this study, one year is divided into 96 time slices: 4 seasons, with each season including 24 times-of-day to create a representative hourly profile for each season. Since one season actually consists of 90 to 92 days, the number of hours in one time slice depends on the number of days in the season where the time slice resides. The segment fraction of a time slice is given by dividing the number of hours in each time slice by the total number of hours in one year.

The demand in an individual time slice (season $s$, time-of-day $d$) is given by averaging the hourly demands during time-of-day $d$ over all days in season $s$. The hourly electricity loads are drawn from FERC Form 714 [15]. Figure S2 compares the modeled load duration curve with the real-world load duration curve for NC. The area under both solid and dashed profiles represents the annual electricity generation, which is the same in both cases. Note the simulated profile does not fully capture the peak load, given the use of 96 versus 8760 (hourly) time slices.

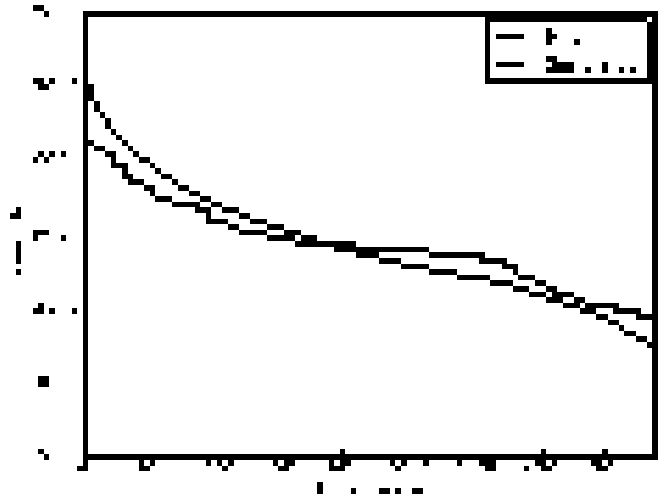

Figure S2. The actual load duration curve (dashed line) and the demand apportioned by model time slice within this study (solid line). The load duration curve is developed using 2014 data from FERC's Form 714 [15].

## 2.1. Technologies and Commodities

There are two categories of energy generation technologies used in the construction of the input dataset for North Carolina: residual (or existing) technologies, and future technologies. Residual technologies represent electricity generating technologies that already have existing capacity in North Carolina. A summary of those technologies and their corresponding Temoa name is presented in Table S4. This data is drawn from EIA's Form 860, which annually reports all electricity generators in each state by prime mover and energy source [16]. Each existing electric sector generator is mapped to a specific Temoa technology, which is described in detail in Table S13.

Within a single technology representation, not all individual generators are the same – older power plants tend to be less efficient and costlier to operate. In addition, as power plants age and reach the end of their useful life, their capacity must be retired. We therefore construct a retirement profile, which reduces the residual capacity of each technology based on an exogenously specified schedule. Temoa is capable of differentiating age within a single technology category using vintages. The vintage represents the year in which capacity for a specific technology was put into service. This differentiation is done by splitting the residual capacity displayed in Table S4 into the appropriate vintage. Using EIA Form 860 data [13], vintages were set up into 5-year bins. For example, capacity built between 1993 and 1997 (inclusive) is categorized as a 1995 vintage. For simplicity, all capacity added prior to 1958 is grouped into the 1960 vintage.

Table S4. Capacity of residual technologies in North Carolina (2015). Data drawn from EIA [10].

| Technology | Description | Capacity (MW) |
|---|---|---|
| EBIOSTMR | Biomass – Steam Turbine | 402 |
| ECOASTMR | Coal – Steam Turbine | 10,803 |
| EDSLCTR | Diesel – Combustion Turbine | 403 |
| EHYDCONR | Hydro – Conventional | 2,004 |
| EHYDREVR | Hydro – Pumped Storage | 86 |
| ELFGGTR | Landfill Gas – Gas Turbine | 15 |
| ELFGICER | Landfill Gas – Integrated Gasification | 60 |
| ENGACCR | Natural Gas – Combined Cycle | 4,766 |
| ENGACTR | Natural Gas – Combustion Turbine | 6,050 |
| ESOLPVR | Centralized Solar PV | 1,437 |
| EURNALWR | Nuclear – Light Water Reactor | 5,114 |
| TOTAL | | 31,140 |

The next step in creating a retirement profile is to specify technology lifetimes. Temoa retires the capacity associated with a given vintage if its lifetime is exceeded at the beginning of an optimization period. EIA provides data on retired power plants across the United States [16], which was used to find the average age of plants that have already been decommissioned. Next, plant lifetime data from the EPA [17], National Renewable Energy Laboratory (NREL) [18], and MiniCam [18] were compared (Table S5). No wind residual technology is specified, as North Carolina did not have any wind capacity as of 2015 [16]. From this data, technology- and vintage-specific lifetimes were chosen for the Temoa model.

Pre-existing, technology-specific capacity split into 5-year vintage bins with the associated technology lifetimes specified for each vintage provides the requisite information needed to model retirements over the 35-year optimization time horizon. Figure S3 presents the retirement profiles for the six largest categories of pre-existing capacity. Note that we assume the lifetime of hydropower plants is 100 years, since many hydropower plants built 50 to 100 years ago are still operating today and upgrades and refurbishment can facilitate generation of carbon-free, cost-effective hydroelectricity. Of the 31.1 GW of pre-existing capacity available to meet demand in 2015, only 3.9 GW remains in 2050. This increasing gap between pre-existing capacity and demand must be met through the addition of new capacity.

Table S5. Comparison of technology lifetimes in years for the US and across energy models [16–18].

| Technology | Avg Lifetime (entire US) | NREL | MiniCAM | EPA MARKAL | This paper |
|---|---|---|---|---|---|
| **EBIOSTMR** | 49 | **45** | **45** | 40 | **45** |
| **ECOASTMR** | 53 | **60** | 45 | 40 | **60** |
| **ECOASTMR_b** | 53 | **60** | 45 | 40 | **60** |
| **EDSLCTR** | 41 | *n/a* | *n/a* | 50 | **45** |
| **EHYDCONR** | 68 | *n/a* | *n/a* | 120 | **100** |
| **EHYDREVR** | *n/a* | *n/a* | *n/a* | 40 | **70** |
| **ELFGGTR** | 16 | **30** | 45 | **30** | 30 |
| **ELFGICER** | 14 | **30** | 45 | 30 | **30** |
| **ENGACCR** | 30 | 30 | 45 | 30 | **40** |
| **ENGACTR** | 37 | 30 | 45 | 30 | **40** |
| **ESOLPVR** | 10 | **30** | **30** | **30** | **30** |
| **EURNALWR** | 32 | **60** | **60** | 40 | **60** |

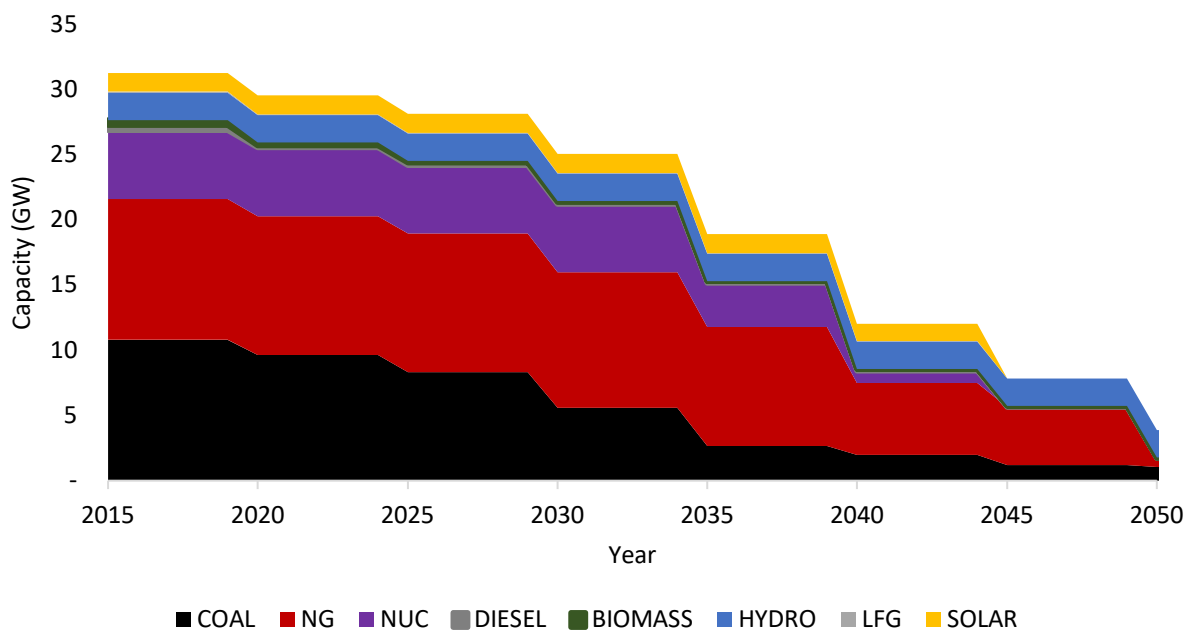

Figure S3. Retirement profile for residual capacity in North Carolina. Note that 'LFG' represents landfill gas, 'NG' represents natural gas, and 'NUC' represents nuclear power.

Table S6. Future energy generation technologies used in NC dataset.

| Technology Name | Description |
|---|---|
| ENGACC05 | Natural Gas – Combined Cycle |
| ENGACT05 | Natural Gas – Combustion Turbine |
| ENGAACC | Natural Gas – Advanced Combined Cycle |
| ENGAACT | Natural Gas – Advanced Combustion Turbine |
| ENGACCCCS | Natural Gas – Combined Cycle with Carbon Capture and Sequestration (CCS) |
| ECOALSTM | Coal – Steam Turbine |
| ECOALIGCC | Coal – Integrated Coal Gasification Combined Cycle |
| ECOALIGCCS | Coal – Integrated Coal Gasification Combined Cycle with CCS |
| ECOALIGCCS_b | Coal – Integrated Coal Gasification Combined Cycle with CCS, baseload |
| ECOALIGCC_b | Coal – Integrated Coal Gasification Combined Cycle, baseload |
| ECOALSTM_b | Coal – Steam Turbine, baseload |
| EURNALWR15 | Nuclear – Light Water Reactors |
| EURNSMR | Nuclear – Small Modular reactors |
| EBIOIGCC | Biomass – Integrated Gasification Combined Cycle |
| ESOLPVCEN | Solar – PV Centralized Generation (Utility scale) |
| ESOLPVDIS | Solar – PV Distributed Generation (Rooftop solar) |
| EWNDON | Wind – Onshore (TRG-9) |
| EWNDOFS | Wind – Shallow Offshore (Generation Class 5) |
| ESLION | Energy storage – Lithium-ion |
| ESCAIR | Energy storage – Compressed air energy storage |
| ESZINC | Energy storage – Zinc-carbon battery |
| ESFLOW | Energy storage – Flow battery |

New capacity is added by Temoa in each period to satisfy electricity demand. For this analysis, new generation technologies were drawn from the EPA MARKAL database [17] and are listed in Table S6. Technologies from this database have detailed technical and cost specifications by US Census Division and provide sufficient representation of electric generation technologies. We also consider an advanced nuclear technology, small modular reactors (SMRs) and four types of utility-scale energy storage technologies: lithium-ion, zinc-carbon, flow batteries and compressed air energy storage systems. Note that although many SMR designs exist [19], we will focus on LWR-based SMR, since it is the most widely used reactor design worldwide and the only reactor type deployed in the US [20]. Each technology used in Temoa represents a specific fuel (e.g., coal or natural gas) and prime mover (e.g., combustion turbine or combined cycle). It is important to note that not all future technologies are currently deployed in North Carolina, but they are made available for future investments. The costs and technical parameters that define each technology determine which technologies are deployed in future periods.

The function of commodities within Temoa is to link technologies together to form a network diagram. There are no technical or cost parameters for commodities. A full list of commodities used is given for reference in Table S14.

## 2.2. Technical Parameters

Each technology in Temoa must be supplied with a set of technical parameters that characterize its operation within each period. The properties in Table S7 broadly define the operational characteristics of each technology in a way that allows the model to meet required physical constraints. Efficiency and emission activity of a technology are linked to specific vintages of a technology, whereas the remaining parameters are the same for all vintages and through all optimization periods. Each of these technical parameters is described in turn.

Table S7. Technical parameters used to define operation of electricity technologies in Temoa.

| Parameter | Description | Source |
|---|---|---|
| Efficiency | The ratio of energy out of a technology to energy in. | 16,17 |
| Availability factor | The maximum amount of electricity that can be produced in a given hourly time slice, relative to nominal capacity. | 17,21,22 |
| Capacity credit | The contribution to peak demand made by non-dispatchable technologies. | 23–25 |
| Emission factors | Kilotons of $CO_2$, $SO_2$, and $NO_X$ emitted per PJ of energy generated. | 26 |
| Baseload classification | Classification of a technology as a "baseload" prevents electricity generation from changing throughout the day to follow varying demand. | 27 |
| Maximum ramp rate | The maximum rate of change (%) of electricity production allowed in a power plant. | 8 |

### 2.2.1. Baseload label

According to the EIA, baseload technologies are designed to satisfy minimum system demand, and "produce electricity at an essentially constant rate and run continuously" [27]. If a technology is classified as baseload, Temoa does not allow the amount of electricity generated to vary hourly within a season. The North Carolina dataset treats nuclear as baseload, per public statements by Duke Energy and the Duke Energy IRPs [6,7,28]. In addition, both existing and future coal are split into two technologies: 40% of the residual capacity is considered baseload per eGRID data [26], while the other 60% is subject to ramp rate constraints. For the new coal capacity added, Temoa will restrict its output variation within each season if it is labeled as baseload, and otherwise it will be subject to ramp rate constraints.

### 2.2.2. Efficiencies

The efficiencies for all the processes are gathered from the EPA MARKAL database [17] and are calibrated with EIA Assumptions to Annual Energy Outlook [29]. Both EPA and EIA provide efficiency estimates for each technology used in Temoa. Because the EPA data is categorized by Census Division, the data for the South Atlantic Division, where North Carolina is located, is used. In addition, operating data drawn from EIA Form 860 [13] and EIA Form 923 [10] is utilized as a more accurate estimate for the efficiencies of all existing power plants. The efficiencies are calculated at the plant level based on the rated capacities reported in EIA Form 860 [13] and the annual electricity fuel consumption reported in EIA Form 923 [10]. The technology-specific efficiency is then given by the weighted average efficiency over all power plants of the same technology. For example, MARKAL estimates that residual natural gas combined-cycle generators in the South Atlantic have an efficiency of 42.2% [17]. However, an analysis of state-level data shows an efficiency of 47.4% [10]. For consistency, state-level data was used to correct the efficiencies of residual technologies. Efficiencies of SMRs are taken from the Westinghouse SMR design [30] given its high burnup and thermal efficiency. Although Westinghouse filed bankruptcy in March 2017 [31], we assume that its technology plans can still be purchased by other vendors. A full listing of all technology efficiencies can be found in Table S19. According to multiple studies [32–35], we assume the round-trip efficiencies of lithium-ion, zinc-carbon, flow batteries and compressed air energy storage systems are 93%, 80%, 65% and 75%, respectively.

### 2.2.3. Availability factors

In Temoa, the availability factor serves as the upper bound on capacity factor. The capacity factor is defined as the ratio of the actual electricity production to the maximum electricity production, if it operated continuously at its full capacity. For dispatchable technologies such as fossil-fuel fired units, the availability factor is set to 90% for all time slices to reflect both forced and unforced outages. For non-dispatchable technologies such as solar and wind power, the availability factors are determined by resource availability.

In this study, the availability factors for all technologies except for wind and solar power are drawn from the EPA MARKAL database. The availability factors of wind and solar power were collected at a higher resolution due to the higher time slice resolution in this study. The data for solar PV is drawn from the NREL Solar Power Data for Integration Studies [21], which consist of 1 year (2006) of 5-minute solar power and hourly day-ahead forecasts for approximately 6,000 simulated PV plants. Note that the plants are categorized into utility scale PV (UPV) and distributed PV (DPV). UPV has single axis tracking while DPV is fixed tilt equal to latitude. In addition, the data for both onshore and offshore wind power is drawn from the NREL Wind Integration National Dataset Toolkit [22]. Note that the availability factors for onshore wind are scaled down such that the annual equivalent capacity factor drops from 38% to 30% because both NREL [36] and EIA [16] show that the capacity factors reported by the WIND Toolkit are approximately 8% to 10% higher than real-world values in most seasons.

To convert these hourly capacity factors to the time slice capacity factors used in Temoa, first the hourly data was categorized by the season and time-of-day slices used in Temoa. The average capacity factor across all sites in North Carolina during a specific season and time of day slice was used in the Temoa dataset. This process resulted in the hourly capacity factor profiles for wind and solar shown in Figure S4, with annual averages shown in Table S8.

Table S8. Average annual capacity factors and capacity credits for solar and wind power.

| Resource Type | Technology | Average Capacity Factor | Capacity Credit |
|---|---|---|---|
| Wind | EWINDON | 30.0% | 20% |
| | EWINDOF | 41.3% | 20% |
| Solar | ESOLPVCEN | 16.9% | 35% |
| | ESOLPVDIS | 15.2% | 35% |
| Energy storage system | ESLION | NA | 70% |
| | ESZINC | | |
| | ESCAIR | | |
| | ESFLOW | | |

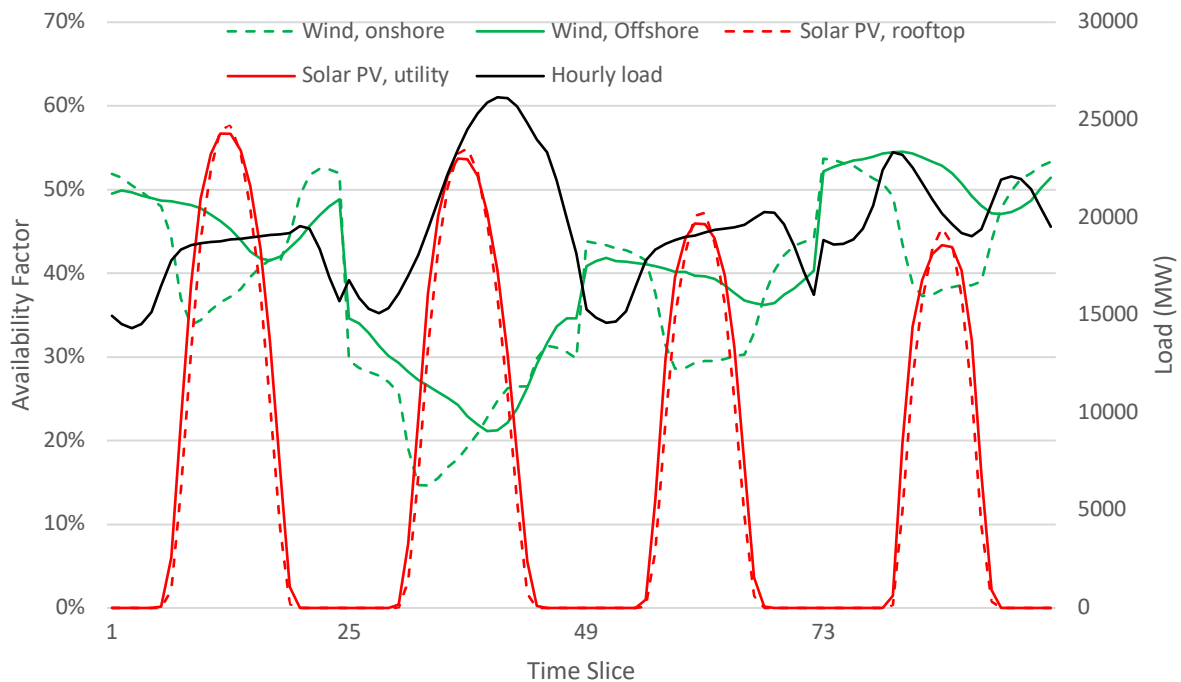

Figure S4. The availability factors for future solar PV and wind power, and the hourly demand in 2015. Time slices are numbered sequentially on the horizontal axis, and the numbers shown correspond to the first timeslice in each season, beginning with spring.

### 2.2.4. Capacity Credits

The hourly capacity factors plotted in Figure S4 indicate that, with some exceptions, wind tends to be most productive during hours of low demand, while solar is most productive during higher demand times. The concept of varying contributions to peak demand relates to capacity credit, which is a measure of how much a resource is able contribute to reducing peak demand [23–25]. Dispatchable generation, such as coal or natural gas, usually has capacity credits close to their nameplate capacity because they can be relied upon to generate during peak periods. By contrast, wind and solar receive less capacity credit because they are not dispatchable during peak demand periods. Methods for estimating the capacity credit of renewable sources differ depending on the metric used [24]. In this study, the capacity credits are utilized in the reserve margin constraints, and we draw capacity credit estimates from a report that utilizes the method of Effective Load Carrying Capability (ELCC). The ELCC of a power generator represents its ability to effectively increase the generating capacity available to a utility or power grid without increasing the utility's loss of load risk [37]. Although previous studies show that capacity credits are affected by factors such as location and existing renewable energy penetration, modeling capacity credits as dependent variables introduces non-linearities into the Temoa formulation. Therefore, it is assumed that capacity credits will remain constant at 20% for onshore [38], 35% for offshore wind [39], and 5% for all forms of solar PV [38], summarized in Table S8. Sioshansi et al. [40] show that capacity credits of electricity storage systems can be affected by hours of storage. Based on their estimates, we use 70% as the capacity credits for all storage technologies with 4 hours of storage. For simplicity, the remaining capacity credits for dispatchable generators are drawn from NERC [41].

### 2.2.5. Emission factors

Emission factors are primarily obtained from the EPA's MARKAL database and eGRID data [26]. This study considers three types of emissions: $NO_X$, $SO_2$, and $CO_2$. Generators in North Carolina were analyzed and cross-referenced to generator details found in EIA Form 860 [13] to calculate technology-specific emission rates. In addition, a network of emission control devices allows Temoa to install carbon capture and sequestration retrofits on coal steam plants, if necessary, to meet the carbon cap. The emission rates used in the NC dataset are summarized in Table S9 for fossil-based generation technologies. Technologies that utilize biomass are considered carbon neutral, as most biomass generation in North Carolina is from wood waste products [10].

We assume that new pulverized coal plants will be equipped with state-of-the-art $SO_2$ and $NO_X$ removal devices and thus no future retrofitting will be required. However, existing coal plants assume uncontrolled emissions of $SO_2$ and $NO_X$, and emission control retrofits are required to meet air quality standards. In this study, $SO_2$ removal through flue gas desulfurization (FGD) is categorized based on coal type and sulfur level, and $NO_X$ removal technologies include low $NO_X$ burners (LNB), selective non-catalytic reduction (SNCR) and selective catalytic reduction (SCR). The capacities of existing retrofit installations are drawn from the EPA eGrid database. In addition, Temoa can install carbon capture and sequestration (CCS) on both existing and new coal units to meet $CO_2$ emission limits.

Table S9. Emission factors of all electricity generating technologies for North Carolina.

| Fossil Fuel Technology | Emission Factors (kt / $PJ_{out}$) | | |
|---|---|---|---|
| | $CO_2$ | $NO_X$ | $SO_2$ |
| ENGACTR | 204 | 0.019 | 0.001 |
| ENGACT05 | 158 | 0.015 | 0.0008 |
| ENGAACT | 126 | 0.012 | 0.0006 |
| ENGACCR | 136 | 0.0128 | 0 |
| ENGACC05 | 100.25 | 0.0094 | 0 |
| ENGAACC | 93.4 | 0.0088 | 0 |
| ENGACCCCS | 11.0 | 0.010 | 0 |
| EDSLCTR | 314.3 | 0.487 | 1.605 |
| EBIOSTMR | 0 | 0.273 | 0.790 |
| EBIOIGCC | 0 | 0.196 | 0.104 |
| ECOALSTM | 227.887 | 0.892 | 3.057 |
| ECOALSTM w/ CCS | 34.124 | 0.892 | 3.057 |
| ECOALIGCC | 33.847 | 0.883 | 3.026 |
| ECOALIGCCS | 10.754 | 0.842 | 2.886 |
| ECOASTMR | 251.193 | 0.983 | 3.369 |

## 2.3. Environmental regulations

North Carolina's emissions of $SO_2$ and $NO_X$ from 1990 to 2014 by source are displayed in Figure S5. Coal contributed over 50% of NC's electricity generation from 1990 to 2007 and is the primary driver of both pollutants. In addition, $NO_X$ emissions from natural gas began to rise after 2010 because more natural gas was employed as a result of its lower price. The $SO_2$ emission factor from natural gas is an order of magnitude lower than from coal [42] and therefore its contribution to $SO_2$ emissions is minimal.

$SO_2$ emissions started to decline in 2005 due to the promulgation of a set of regulations at both the federal [43] and state level [44] under the Clean Air Act. The EPA's Clean Air Interstate Rule (CAIR), a cap-and-trade program intended to reduce $SO_2$ and $NO_X$ emissions beyond the levels defined by the Acid Rain Program in the eastern half of the United States, led to the retrofit of 91 GW [45] of coal-fired power capacity with FGD scrubbers between 2005 and 2011 nationwide. In addition, in 2002 North Carolina implemented the Clean Smokestacks Act, its own state-level regulation [44], resulting in a significantly faster reduction of $SO_2$ emissions than in neighboring states [46] between 2002 and 2012. Likewise, emissions of $NO_X$ peaked in 1997 at 280 kilotons and declined considerably thereafter due to the implementation of the Acid Rain Program, whose first phase spanned from 1996 to 2000 [47] and the Clean Smokestacks Act [44]. To comply with the regulations, utilities in NC begin by adding $NO_X$ control technologies (before 2007), including SCR and SNCR, and switch to replacing older units with units equipped with state of the art control technology after 2007 due to higher than anticipated retrofit cost [48].

In 2011, the US EPA finalized the Cross-State Air Pollution Rule (CSAPR) as a replacement to the CAIR program, requiring 28 states in the eastern United States to reduce $SO_2$, annual $NO_X$, and ozone season $NO_X$ emissions from fossil fuel-fired power plants [49]. North Carolina is among the 28 affected states that reduce its annual $SO_2$ and $NO_X$ emissions. CSAPR sets an "assurance limit" in 2015 and 2017 for each state, which takes into account their historical production and available control technologies [50]. This assurance limit consists of a "budget" (primary allocation) and "variability" (provided for flexibility). To model this policy in Temoa, we apply the 2015 assurance limit in the first period, and the 2017 budget limit in all later periods, as we assume no policy changes in our baseline dataset (and thus no additional regulation of $SO_2$ and $NO_x$ limits beyond CASPR limits). The CASPR limits are displayed in Figure S5.

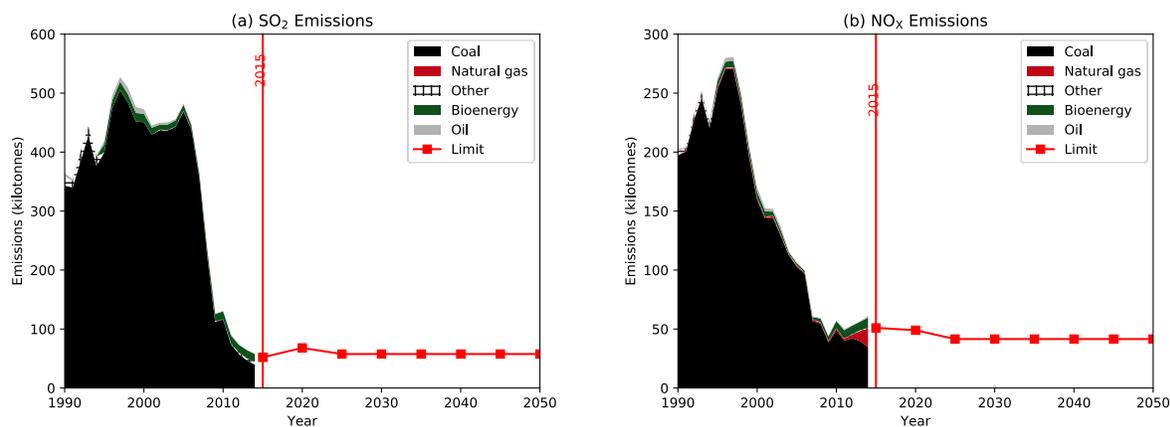

Figure S5. Historical emissions and emission limits of (a) $SO_2$ and (b) $NO_X$ for North Carolina's electric power sector through 2050. Emission sources are distinguished by fuel type. The vertical red line indicates the first period of the optimized horizon, which is 2015. The red squares after 2015 represent existing emission limits, which are assumed to remain place for the duration of the model time horizon.

Beyond the emission control regulations, North Carolina implemented a Renewable Energy Portfolio Standard in 2007 [51]. This legislation sets requirements for utilities to generate a certain percentage of their electricity sales from renewable sources. The REPS targets are converted from percentages into electricity generation requirements and enforced by implementing a minimum generation constraint over each optimization period. These requirements are enumerated in Table S10 below. The RPS is implemented in Temoa via a minimum activity constraint [1]. This is handled by creating an intermediary technology which converts the commodity "renewable electricity" into "electricity" at 100% efficiency and zero cost. Each technology in Temoa that is covered by the RPS (i.e., solar, geothermal, wind, biomass, landfill gas, and small hydro) is modified to generate "renewable electricity" instead of "electricity". The minimum activity constraint is then applied to this new intermediary technology in each Temoa optimization period. Minimum required shares for swine and poultry waste are ignored in this analysis, as they make up less than one percent of total electricity demand, and reliable cost data for these technologies are not publicly available at this time.

Table S10: Renewable Energy Portfolio Standard in North Carolina. Data drawn from NC Utilities Commission [51].

| Calendar Year | RPS Requirement | Carveouts | | | Maximum share of RPS Allowed from Energy Efficiency |
| | | Solar Energy | Swine Waste | Poultry Waste (MWh) [a] | |
|---|---|---|---|---|---|
| 2012 | 3% | 0.07% | 0.07% | 170,000 | 25% |
| 2015 | 6% | 0.14% | 0.14% | 900,000 | 25% |
| 2018 | 10% | 0.20% | 0.20% | 900,000 | 25% |
| 2021 and thereafter | 12.5% | 0.20% | 0.20% | 900,000 | 40% |

a. Compliance through the use of poultry waste is specified in MWh given its small percentage share.

This study also accounts for the federal investment tax credit (ITC). The federal ITC applies to residential and business investments [52] of eligible technologies: investment costs of all types of solar PV and wind power are reduced by 30% before 2020; investments of all solar PV are reduced by 26% before 2022, and a 10% reduction is also applied to utility scale solar PV thereafter.

## 3. Costs

Temoa minimizes the total present cost, which has three components: investment costs, fixed costs, and variable costs. Investment costs represent the initial capital outlay plus loan costs needed to build new capacity, fixed costs represent operations and maintenance costs that are independent of generation level, and variable costs include operational expenses that are dependent on the generation level. Variable costs are also used to specify fuel commodity prices. Cost estimates were obtained primarily from the Environmental Protection Agency's (EPA) MARKAL database [17] and NREL's Annual Technology Baseline (ATB) 2018 [53], and are described below in more detail.

Capital costs for new generating technologies are drawn from ATB 2018, which includes the overnight capital cost, grid connection cost, and construction financing cost for each technology. For conventional generating technologies, such as coal and natural gas, the cost estimates from ATB 2018 are less uncertain due to their commercial maturity. Therefore, the costs from the "Mid Technology Cost Scenario" are used in this analysis. Due to the rapid drop in the solar PV capital cost, the ATB 2018 cost was cross-checked with a variety of other sources, including EIA's Assumptions to the AEO 2017 [29] and Fu et al. [54]. In ATB 2018, solar PV costs are categorized by capacity factors. According to NREL's Solar Grid Integration Study [21], the annual capacity factor of NC's utility-scale solar PV is 16.9%, therefore, the cost group of "utility-14% mid" cost projection is selected; similarly, the mid-cost projection is selected for distributed residential solar PV. Note that capital costs of solar PV are reported in $/kW_{dc} in the ATB data and converted into kW_{ac} using a conversion factor of 1.2 kW_{ac}/kW_{dc}. NREL categorizes land-based wind into 10 techno-resource groups (TRGs) based on wind speed. According to NREL's annual average wind speed map [55], most of NC's territory has a wind speed between 4.0 to 4.5 m/s at 100 m hub height, therefore, capital costs from TRG 9 in ATB 2018 are selected. In addition, the federal investment tax credits for renewable energy [52] are accounted for by reducing the associated investment costs. For example, the federal business tax credit provides a 30% tax credit

for qualified renewable expenses between 2016 and 2019, which is reduced to 10% in all years after 2022. This discount was applied directly to the investment costs for the technologies covered under the legislation [52]. While the SMR capital cost is subject to great uncertainty due to technology immaturity and a lack of commercial operating experience, Carelli et al. [56] find that the ratio of unit capital cost of SMR to LWR ranges from 1.05 to 1.16, and an expert elicitation conducted by Abdulla et al. [57] shows the SMR units with lower capacities are more expensive in terms of overnight cost. Therefore, we assume the overnight SMR cost is 5% higher than the LWR estimate, since the rated capacity of the Westinghouse SMR is at the high end of three proposed technologies in the United States. In addition, we assume a 3-year lead time for SMR construction [56], compared to 6 years for the LWR, in order to calculate the financial costs incurred. We further assume that the discount rate of SMRs and LWRs are both 15%, as opposed to 10% for other technologies, to reflect higher financial risk associated with investment in nuclear power plants. A full listing of new technologies and their investment costs can be found in Table S15.

Investment costs in pollution control retrofits as well as fixed and variable operations and maintenance costs are drawn from the EPA nine-region MARKAL database [17]. The EPA maintains its own MARKAL-compatible dataset of energy costs based on the nine Census Divisions. For the purposes of this model, costs from the South Atlantic region were used for North Carolina. Pollution control capital costs in the MARKAL dataset are in units of million dollars per gigawatt (M$/GW), in 2005 dollars. The costs used in this North Carolina dataset are 2015 constant dollars, so the cost estimates from the EPA were adjusted using the Consumer Price Index [58]. Fixed costs are incurred annually and specified in units of million dollars per gigawatt-year (M$/GW-yr), while variable costs are proportional to electricity generated and specified in units of million dollars per petajoule (M$/PJ). Most variable and fixed costs are drawn from the MARKAL database following the same methodology as discussed above. A complete listing of fixed and variable O&M costs can be found in Table S16–Table S18.

Costs for electricity storage technologies depend on technical characteristics such as energy storage capacity and charging-discharging power capacity. Therefore, capital costs of electricity storage systems are typically defined in terms of power capacity ($/kW) or energy capacity ($/kWh). In this study, we assume a fixed 4-hour storage duration, so the capital costs of storage technologies are expressed in terms of power capacity (M\$/GW) to be consistent with all generating technologies. Capital and fixed O&M costs for lithium-ion batteries are based on GTM Research [59] and cost data for the other storage technologies are taken from Zakeri and Syri [60]. All storage costs are included in Table S15–Table S18. Note we assume zero variable O&M costs for all storage technologies.

The costs and technical parameters described above were used to create the baseline dataset for North Carolina. Scenario-specific parameters include fossil fuel costs and the upper bound placed on $CO_2$ emissions.

## 4. Break-even analysis based on reduced cost

### 4.1. Derivation

Consider the following standard form linear program:

$$\min c^T x \tag{3a}$$
$$\text{s.t. } Ax = b \tag{3b}$$
$$x \geq 0 \tag{3c}$$

Suppose the constraint matrix $A$ can be decomposed into $[B, N]$, where $B$ represents the columns of matrix $A$ associated with the basic variables at a particular feasible solution (this matrix is of full rank) and $N$ represents the columns of matrix $A$ associated with the non-basic variables. Analogously, the vector of decision variables denoted by $x$ can be decomposed into $x = [x_B^T, x_N^T]^T$, where $x_B$ and $x_N$ represent the decision variables associated with $B$ (i.e., basic variables) and $N$ (i.e., non-basic variables), respectively. Similarly, the vector of objective function coefficients $c$ can be decomposed in a similar way, i.e. $c = [c_B^T, c_N^T]^T$.

When model (3a)–(3c) is solved and an optimal solution is found, the objective function becomes:

$$c^T x^* = c_B^T B^{-1} b + (c_N^T - c_B^T B^{-1} N) x_N^* \tag{4}$$

where, $x^*$ represents the optimal decision vector; the reduced cost vector $r_N$ associated with non-basic variables $x_N^*$, which are the coefficients of $x_N^*$, should be greater than or equal to zero:

$$r_N^T = c_N^T - c_B^T B^{-1} N \geq 0 \tag{5}$$

Under optimality conditions, the non-basic variables $x_N^* = 0$ correspond to the technologies that are not deployed in Temoa. Therefore, if we want to make one specific non-basic variable $x_j$ enter the basis, i.e., be part of the solution and included in the basic variables vector, its reduced cost must become negative, all else equal (i.e., $c_B^T B^{-1} N$ remains constant). Therefore, its coefficient must be reduced to $(c_j - r_j)$, where $c_j$ and $r_j$ are the objective function coefficient and the reduced cost associated with entering variable $x_j$, respectively.

In Temoa's objective function, the reduced cost is associated with the capacity variable $CAP_{t,v}$, which is the installed capacity associated with vintage $v$ of technology $t$. Note that the full notation below is explained in Hunter et al. [1]. There are two separate objective function terms associated with $CAP_{t,v}$, which involve the calculation of loan costs $C_{loans}$ and fixed operations and maintenance costs $C_{fixed}$:

$$C_{loans} = \sum_{t,v \in \Theta_{IC}} \left( \left[ IC_{t,v} \cdot LA_{t,v} \cdot \frac{(1+GDR)^{P_0-v+1} \cdot (1-(1+GDR)^{-LLN_{t,v}})}{GDR} \cdot \frac{1-(1+GDR)^{-LPA_{t,v}}}{1-(1+GDR)^{-LP_{t,v}}} \right] \cdot CAP_{t,v} \right)$$

$$C_{fixed} = \sum_{p,t,v \in \Theta_{IC}} \left( \left[ FC_{p,t,v} \cdot \frac{(1+GDR)^{P_0-p+1} \cdot (1-(1+GDR)^{-MPL_{t,v}})}{GDR} \right] \cdot CAP_{t,v} \right)$$

The coefficient of decision variable $CAP_{t,v}$ therefore involves the constant terms from $C_{loans}$ and $C_{fixed}$:

$$c_j = IC_{t,v} \cdot LA_{t,v} \cdot \frac{(1+GDR)^{P_0-v+1} \cdot (1-(1+GDR)^{-LLN_{t,v}})}{GDR} \cdot \frac{1-(1+GDR)^{-LPA_{t,v}}}{1-(1+GDR)^{-LP_{t,v}}}$$
$$+ \sum_p FC_{p,t,v} \cdot \frac{(1+GDR)^{P_0-p+1} \cdot (1-(1+GDR)^{-MPL_{t,v}})}{GDR}$$

where $j$ denotes the column index of $CAP_{t,v}$ in the vector of decision variables. $c_j$ is a constant term and can be expressed as a linear combination of capital cost $IC_{t,v}$ and fixed O&M cost $FC_{p,t,v}$:

$$c_j = \alpha \cdot IC_{t,v} + \sum_p \beta_p \cdot FC_{p,t,v}$$

where $\alpha$ and $\beta_p$ represent the remaining constant terms in $c_j$; note that the fixed cost is indexed by time period $p$. Once the solver returns the reduced cost $r_j$ associated with $CAP_{t,v}$, the break-even capital cost $IC_{t,v}^{BE}$ is given by the following equation, assuming $\alpha$, $FC_{p,t,v}$, and $\beta_p$ remain constant:

$$IC_{t,v}^{BE} = \frac{1}{\alpha}\left(c_j - r_j - \sum_p \beta_p \cdot FC_{p,t,v}\right) = IC_{t,v} - \frac{r_j}{\alpha} \tag{6}$$

## 4.2. Implementation

In practice, we initially encountered some difficulty with this approach. Figure S6 shows installed capacities (green line) of biomass IGCC in the Cap-H scenario as a function of its investment cost in 2020, which is manually varied with a cost multiplier. In this case, the original investment cost of biomass IGCC in 2020 is 3,658 $/kW, but it is not deployed in 2020 until its investment cost drops below 2,158 $/kW, which represents 59% of its original value. By definition, the break-even cost of biomass IGCC in 2020 is 2,158 $/kW. The break-even costs are represented as a function of capital cost multipliers in Figure S6 (dashed red line). When the cost multiplier is greater than 59%, the break-even cost is always 2,158 $/kW, as represented by the horizontal part of the line. By contrast, when the cost multiplier is less than 59%, the break-even cost is simply the product of the cost multiplier and 3,658 $/kW.

Figure S6 also includes estimates derived from the reduced costs returned by CPLEX with three algorithms: barrier, primal simplex and dual simplex. The primal and dual simplex estimates are higher than the actual values when cost multipliers are greater than 63%, while estimates from all three algorithms are lower than the actual values when cost multipliers are between 51% and 59%. Therefore, the reduced cost estimates occasionally fail to return the correct break-even costs. In summary, estimates based on reduced costs tend to overestimate the break-even costs if the technology under consideration is not deployed, and to underestimate the break-even costs if the technology has been deployed.

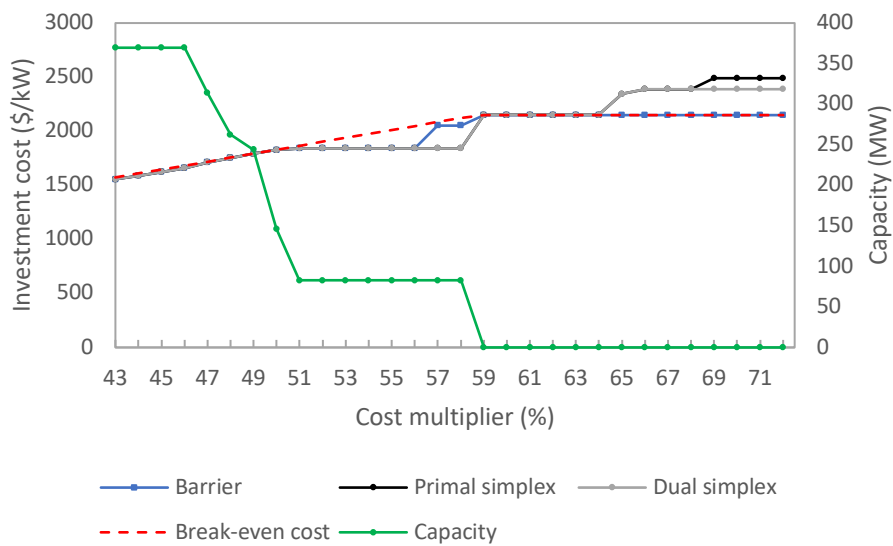

Figure S6. Investment cost and installed capacity as a function of the cost multiplier. The green line represents the installed capacity of biomass IGCC in 2020 as a function of the cost multiplier, the red line represents the realized break-even investment cost as a function of the cost multiplier, and the blue, black, and gray lines represent break-even costs calculated from reduced costs as a function of the cost multiplier returned by three different solution algorithms employed by CPLEX. Ideally, the break-even costs obtained through reduced costs should match the red line.

Although errors in the break-even cost estimates may be incurred due to imperfections in this approach, several methods can be employed to improve the estimates. First, if a technology is deployed during a specific period, the break-even cost is known, since it must equal its investment cost in that period due to complementary slackness, which is a technique used to verify that a solution is optimal [61]. Therefore, as shown in Figure S6, the break-even costs are determined immediately when the cost multiplier is smaller than 59%. In addition, as Figure S6 suggests, when a technology is not deployed, the correct break-even costs can be obtained by selecting the minimum estimate derived from reduced costs returned by different algorithms, or different solvers.

A binary search method can also be employed to derive the correct break-even costs of a specific technology according to the definition of break-even cost. However, the brute force binary search method is significantly more time consuming than the reduced cost approach outlined in this study, since each complete run of binary search includes an indefinite number of independent model solves, as opposed to only one single run when using reduced costs. Furthermore, each run of the binary search only returns the break-even cost of one technology in a single year, while the reduced cost approach can return break-even costs of all technologies in every period. The results presented in this paper are derived using multiple solvers and different algorithms and have been calibrated with solutions from the binary search method. Our results suggest that the reduced costs based method can return the correct break-even costs after applying the complementary slackness condition.

## 4.3. Comparison with levelized cost

A commonly used metric to compare technology cost-effectiveness is the levelized cost of electricity (LCOE), which can be expressed as the ratio of annualized lifetime cost of a technology over its annual electricity production. In our analysis, the annualized cost accounts for investment cost, operations and maintenance cost, and fuel costs. For deployed technologies, the total annual electricity production by deployed technologies is directly given by the model solution. For technologies that are not deployed, their annual electricity production is estimated using their availability factors (detailed discussion given in Section 2.2.3). Therefore, the technology-specific LCOEs associated with technologies that are not deployed are only an approximation.

We use the grid LCOE as the baseline to conduct the comparative study. Similar to the technology-specific LCOE, the grid LCOE can be expressed as the ratio of the total annual cost of the entire electric system over the sum of annual electricity production from all technologies. As mentioned in the introduction, a drawback associated with LCOE is its failure to account for grid dynamics. We demonstrate the limitation of LCOE comparisons using two examples described below.

In the first example, we examine the LCOE of single cycle natural gas combustion turbines (NGCT) in the L scenario. This technology is deployed in the L scenario, and the capacity factors are directly drawn from the model output. As shown in Table S11, the LCOE of NGCT ranges from 500 to over 3,000 $/MWh, which is prohibitively high compared with the grid LCOE (presented in Figure S7). However, the LCOE fails to capture the contribution of NGCT towards the reserve margin constraint: NGCT is one of the cheapest dispatchable technologies given its low investment cost, and most time it remains idle to meet the reserve margin constraint described in Section 1.1 above. Therefore, the high LCOE is due primarily to its low

utilization rate, which fails to reflect its value as reserve capacity.

Table S11. LCOE of single cycle natural gas combustion turbine in the L scenario.

| Period | Capacity factor | LCOE ($/MWh) |
|--------|-----------------|--------------|
| 2030 | 0.29% | 3,488 |
| 2035 | 0.67% | 1,520 |
| 2040 | 0.34% | 2,955 |
| 2045 | 1.09% | 930 |
| 2050 | 1.94% | 531 |

The second example pertains to onshore wind, which is not deployed under most scenarios. Figure S7 shows that the LCOE of onshore wind power is greater than the grid LCOE across all time periods and scenarios, except for the 2050 period in the Cap-H scenario. Therefore, comparison of LCOE implies that onshore wind is cost-competitive only in the last period of the Cap-H scenario. However, Figure S11 shows that onshore wind has been deployed in time periods 2040, 2045, and 2050 in the Cap-H scenario. Thus, the prediction based on the LCOE comparison in Figure S7 is inaccurate, while the break-even analysis correctly reflects the true cost-effectiveness of onshore wind in the model from 2040 to 2050, as shown in Figure 3. We speculate that wind has additional value when accounting for the system constraints under the emissions cap, which is not captured by simple LCOE comparisons.

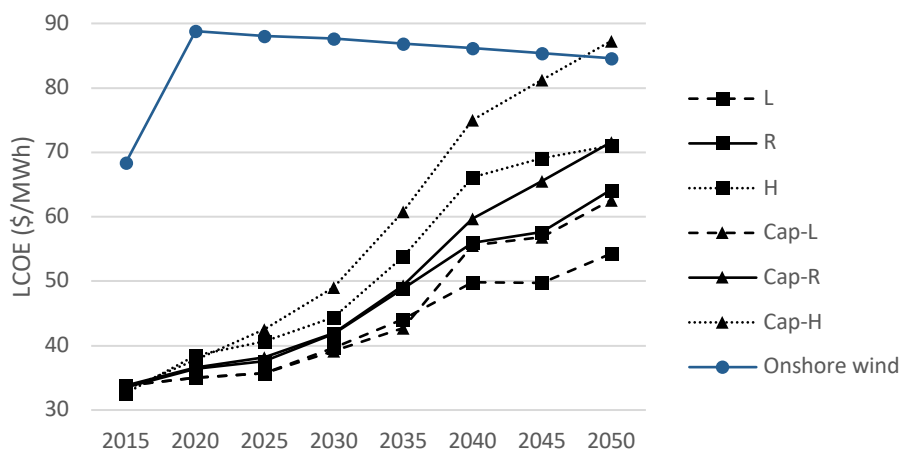

Figure S7. Comparison of onshore wind and grid LCOE from select scenarios. The increasing trends of grid LCOE are due to the aggregation of loans from newly invested units. Note that transmission and distribution costs are not included.

# 5. Supplementary tables and figures

In addition to the above data for the baseline scenario, our study also considers uncertainties associated with fuel prices, a hypothetical $CO_2$ cap, and an extended renewable portfolio standard. Table S12 summarizes all scenarios in this study. Figure S8 shows natural gas and coal prices from all three fuel price scenarios. Figure S9 illustrates NC's historical $CO_2$ emissions along with the hypothetical carbon cap, which is based on the climate action plan of NC's largest investor-owned utility but linearly extrapolated to 2050. The mass-based emissions target associated with the EPA Clean Power Plan (CPP) is included for reference. Figure S10 illustrates the extended REPS, including the assume breakdown between energy efficiency, renewable energy, and non-renewable energy. Figure S11 includes the installed capacity across the tested scenarios, and represents the analogue to the electricity generation plotted in Figure 1 in the manuscript.

Table S12. Scenarios and the associated labels used in this study.

| Scenario | Natural gas prices | Carbon cap | Policy |
|----------|-------------------|------------|--------|
| L | Low | No | |
| R | Reference | No | |
| H | High | No | |
| Cap-L | Low | Yes | |
| Cap-R | Reference | Yes | |
| Cap-H | High | Yes | |
| REPS-R | Reference | No | Extended REPS |
| REPS-R-ITC | Reference | No | Extended REPS and 40% ITC for onshore wind |

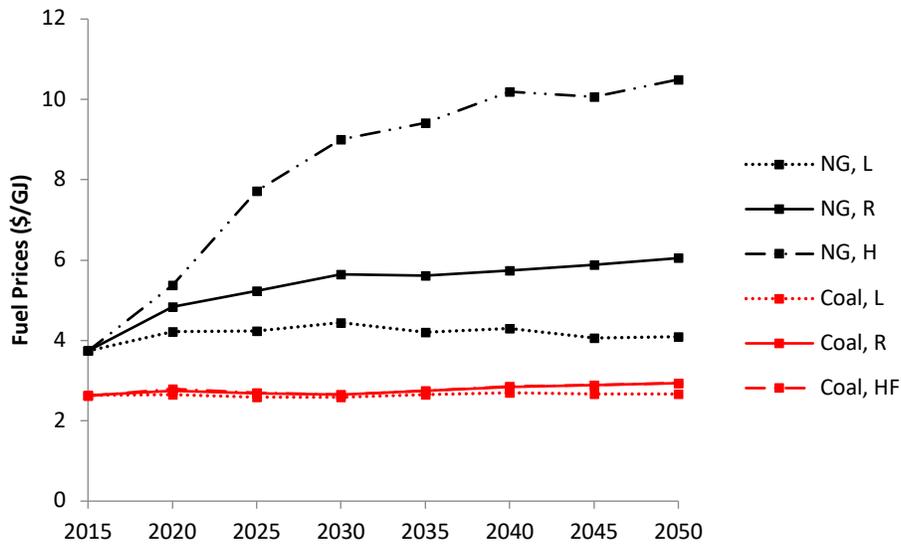

Figure S8. Forecasted prices of natural gas and coal. Note that the fuel prices in the R, L and H scenarios are drawn from the Reference, High Oil and Gas Resource, and Low Oil and Gas Resource scenarios in AEO2017, respectively.

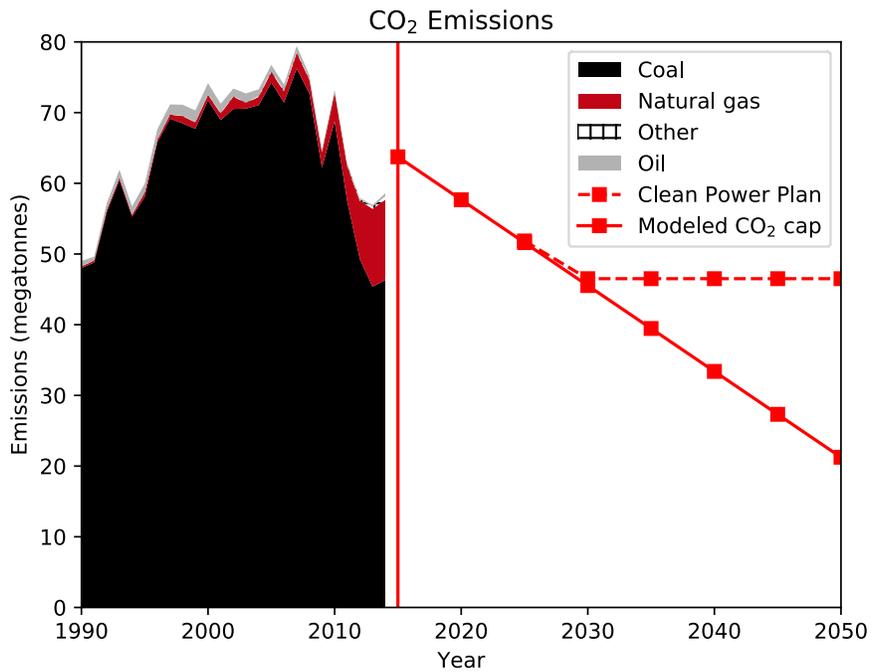

Figure S9. Historical $CO_2$ emissions from North Carolina's electric power sector, the modeled carbon cap [62], and for reference, the mass-based emissions targets associated with EPA Clean Power Plan.

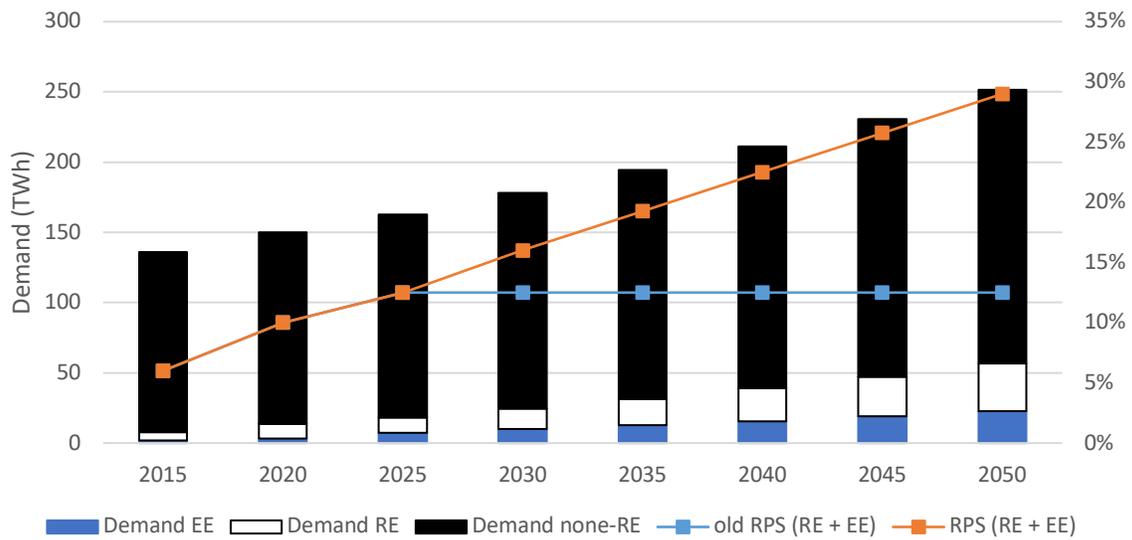

Figure S10. The extended REPS. The bars represent the break-down of demand by categories: demand served by renewable energy (RE), energy efficiency (EE) and everything else, and the lines represent required fractions of energy from renewable energy and energy efficiency measures.

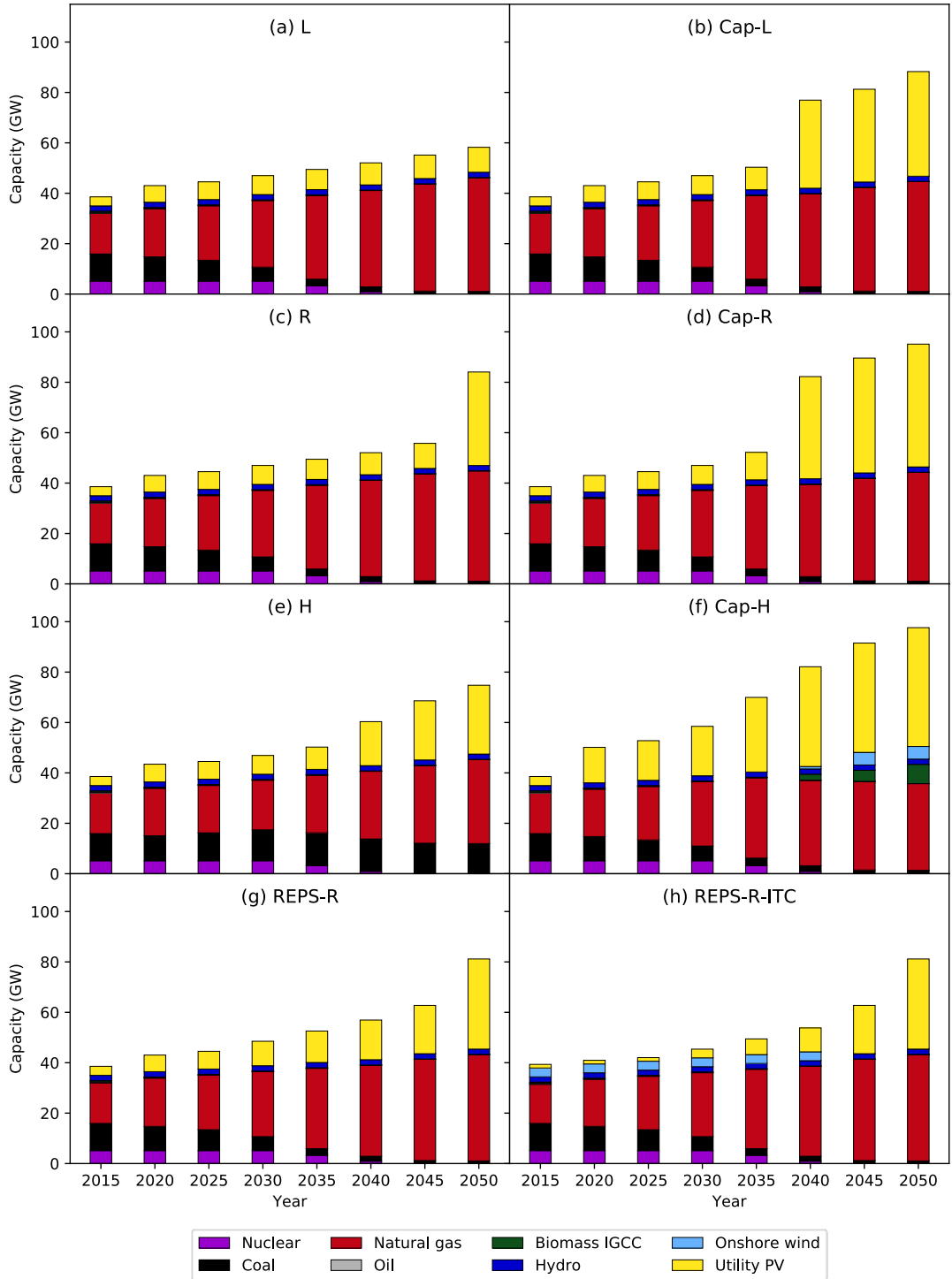

Figure S11. The electricity capacity mixes of North Carolina through 2050 for all eight scenarios. (a) low natural gas prices [L], (b) carbon cap with low natural gas prices [Cap-L], (c) reference natural gas prices [R], (d) carbon cap with reference natural gas prices [Cap-R], (e) high natural gas prices [H], (f) carbon cap with high natural gas prices [Cap-H], (g) extended REPS with reference natural gas prices [REPS-R], and (h) extended REPS with reference natural gas prices and an investment tax credit for onshore wind [REPS-R-ITC].

# 6. Datasets

## 6.1. Dataset A – Technologies

Table S13. Technologies mapped from EIA Form 860 to this study.

| Technology | Prime Mover [a] | Energy Source [b] | TEMOA Tech [c] | Existing Capacity (MW) [d] |
|---|---|---|---|---|
| All Other | ST | TDF | unknown | 107.0 |
| All Other | ST | WH | unknown | 54.0 |
| Conventional Hydroelectric | HY | WAT | EHYDCONR | 2,004.1 |
| Conventional Steam Coal | ST | BIT | ECOASTMR | 10,802.8 |
| Hydroelectric Pumped Storage | PS | WAT | EHYDREVR | 86.0 |
| Landfill Gas | FC | LFG | unknown | 10.0 |
| Landfill Gas | GT | LFG | ELFGGTR | 15.4 |
| Landfill Gas | IC | LFG | ELFGICER | 59.9 |
| Natural Gas Combined Cycle | CA | NG | ENGACCR | 1,812.8 |
| Natural Gas Combined Cycle | CT | NG | ENGACCR | 2,953.2 |
| Natural Gas Combustion Turbine | GT | NG | ENGACTR | 6,049.7 |
| Nuclear | ST | NUC | EURNALWR | 5,113.6 |
| Other Waste Biomass | ST | SLW | EBIOSTMR | 0.8 |
| Petroleum Liquids | GT | DFO | EDSLCTR | 241.0 |
| Petroleum Liquids | IC | DFO | EDSLCTR | 161.8 |
| Solar Photovoltaic | PV | SUN | ESOLPVR | 1,436.8 |
| Wood/Wood Waste Biomass | ST | WDS | EBIOSTMR | 209.7 |
| Wood/Wood Waste Biomass | ST | BLQ | EBIOSTMR | 191.7 |

a. Prime mover code from EIA Form 860: ST – Steam turbine, HY – Hydro turbine, PS – Energy storage, FC – Fuel cell, GT – Gas turbine, IC – Internal combustion engine, CA – Combined cycle steam part, CT – Combined cycle combustion turbine part, PV – Photovoltaic.

b. Energy source code from EIA Form 860. Note that although two fuel sources are provided for some technologies, the technology in EIA Form 860 is mapped to Temoa technology only based on the type of prime mover and energy source 1. Energy source codes: TDF – Tire-derived fuels, WH – Waste heat, WAT – Water, BIT – Bituminous coal, LFG – Landfill gas, NG – Natural gas, NUC – Nuclear, SLW – Sludge waste, DFO – Distilled fuel oil (including diesel, No. 1, No. 2, and No. 4 fuel oils), SUN – Solar, WDS – Wood/Wood waste solid, BLQ – Black liquor.

c. Technologies named "unknown" are excluded.

d. Note that summer capacities from EIA Form 860 are used to calculate existing capacities following traditions in EIA State Electricity Profiles.

## 6.2. Dataset B – Commodities

Table S14. List of commodities used in this study.

| Commodity | Sector<br>p = physical<br>e = emissions | Description |
|---|---|---|
| Ethos | p | Dummy commodity to supply inputs |
| COALSTMCC | p | Coal |
| COALIGCCCC | p | Coal |
| COALIGCC | p | Coal |
| COALSTM | p | Coal |
| ELCNGAEA | p | Natural Gas |
| ELCDSLEA | p | Diesel |
| LFGICEEA | p | Landfill gas to ICE |
| LFGGTREA | p | Landfill gas to gas turbines |
| URNA | p | Uranium |
| ELCBIGCCEA | p | Biomass to IGCC |
| ELCBIOSTM | p | Biomass to steam |
| ELCGEO | p | Geothermal |
| SOL | p | Solar |
| WND | p | Wind |
| ELCHYD | p | Hydro |
| ELCRNWB | p | Electricity, physical, from renewables |
| ELC | p | Electricity, physical, to transmission |
| ELCDIS | p | Electricity, physical, to distribution |
| ELCDMD | d | Electricity, demand |
| co2 | e | $CO_2$ emissions |
| so2_ELC | e | $SO_2$ emissions from the electric sector |
| nox_ELC | e | $NO_X$ emissions from the electric sector |
| so2_SUP | e | $SO_2$ emissions from the supply sector |
| nox_SUP | e | $NO_X$ emissions from the supply sector |
| COALSTM_R_B | p | Existing BIT coal steam to the blending tech |
| COAB_R | p | Existing BIT coal after SCR/SNCR or SCR PT to the bit blending technology for existing coal steam |
| COAB_R_SCR_PT | p | Existing bituminous coal after LNB retrofit or passthrough to the SCR SNCSR $NO_X$ retrofit or passthrough |
| COAB_R_LNB | p | Existing bituminous coal after $CO_2$ capture to the LNB retrofit |
| COAB_R_LNB_PT | p | Existing bituminous coal after $SO_2$ or $CO_2$ passthrough to the LNB nox retrofit or passthrough |
| COAB_R_CC | p | Existing bituminous coal after $SO_2$ removal to the $CO_2$ capture retrofit or passthrough |

## 6.3. Dataset C – Costs

Table S15. List of overnight investment costs in M$/GW or $/kW indexed by technology and vintage. Data sources used for cost projections are discussed in Section 3.

| Technology | 2015 | 2020 | 2025 | 2030 | 2035 | 2040 | 2045 | 2050 |
|---|---|---|---|---|---|---|---|---|
| EBIOIGCC | 3,805.0 | 3,657.0 | 3,582.0 | 3,582.0 | 3,508.0 | 3,508.0 | 3,508.0 | 3,508.0 |
| ECOALIGCC | 4,240.8 | 4,180.1 | 4,068.4 | 3,954.5 | 3,862.9 | 3,774.0 | 3,687.5 | 3,575.2 |
| ECOALIGCCS | 6,494.0 | 6,121.0 | 5,747.0 | 5,747.0 | 5,747.0 | 5,747.0 | 5,747.0 | 5,747.0 |
| ECOALSTM | 3,952.1 | 3,902.6 | 3,849.8 | 3,802.0 | 3,754.8 | 3,704.3 | 3,658.7 | 3,578.1 |
| EGEOBCFS | 2,517.0 | 2,391.0 | 2,328.0 | 2,328.0 | 2,266.0 | 2,266.0 | 2,266.0 | 2,266.0 |
| ENGAACC | 1,054.5 | 1,048.4 | 1,024.7 | 1,001.1 | 983.4 | 966.4 | 950.1 | 927.1 |
| ENGAACT | 901.7 | 896.0 | 874.1 | 852.1 | 836.1 | 821.0 | 806.3 | 786.5 |
| ENGACC05 | 923.0 | 923.0 | 923.0 | 923.0 | 923.0 | 923.0 | 923.0 | 923.0 |
| ENGACCCCS | 2,201.0 | 2,167.6 | 2,079.3 | 1,990.8 | 1,917.6 | 1,847.1 | 1,777.1 | 1,696.9 |
| ENGACT05 | 979.0 | 979.0 | 979.0 | 979.0 | 979.0 | 979.0 | 979.0 | 979.0 |
| ESOLPVCEN | 1,618.0 | 927.6 | 1,021.1 | 960.7 | 919.2 | 877.7 | 830.2 | 782.8 |
| ESOLPVDIS | 3,441.7 | 2,218.2 | 2,362.0 | 1,941.4 | 1,776.7 | 1,612.0 | 1,547.0 | 1,482.0 |
| ESOLSTCEN | 2,788.0 | 2,758.0 | 3,417.0 | 3,337.0 | 3,178.0 | 3,178.0 | 3,178.0 | 3,178.0 |
| EURNALWR15 | 6,157.9 | 6,143.4 | 6,043.7 | 5,886.8 | 5,730.7 | 5,569.6 | 5,415.8 | 5,210.7 |
| EURNSMR | 5,469.0 | 5,196.0 | 5,060.0 | 5,060.0 | 4,922.0 | 4,922.0 | 4,922.0 | 4,922.0 |
| EWNDOFS | 3,843.7 | 5,216.0 | 4,942.0 | 4,667.0 | 4,393.0 | 4,393.0 | 4,393.0 | 4,393.0 |
| EWNDON | 1,214.3 | 1,738.4 | 1,738.4 | 1,738.4 | 1,738.4 | 1,738.4 | 1,738.4 | 1,738.4 |
| ESLION | 2,370.0 | 1,741.0 | 1,446.0 | 1,308.0 | 1,212.0 | 1,139.0 | 1,107.0 | 1,078.0 |
| ESZINC | 2,160.0 | 1,987.0 | 1,868.0 | 1,756.0 | 1,651.0 | 1,568.0 | 1,505.0 | 1,460.0 |
| ESFLOW | 2,976.0 | 2,143.0 | 1,309.0 | 1,232.0 | 1,155.0 | 1,110.0 | 1,065.0 | 971.0 |
| ESCAIR | 1,576.0 | 1,497.0 | 1,422.0 | 1,351.0 | 1,297.0 | 1,258.0 | 1,233.0 | 1,221.0 |
| E_LNBSCR_COAB_N | 1.535 | 1.535 | 1.535 | 1.535 | 1.535 | 1.535 | 1.535 | 1.535 |
| E_LNBSNCR_COAB_N | 0.786 | 0.786 | 0.786 | 0.786 | 0.786 | 0.786 | 0.786 | 0.786 |
| E_SNCR_COAB_N | 0.544 | 0.544 | 0.544 | 0.544 | 0.544 | 0.544 | 0.544 | 0.544 |
| E_SCR_COAB_N | 1.284 | 1.284 | 1.284 | 1.284 | 1.284 | 1.284 | 1.284 | 1.284 |
| E_LNB_COAB_N | 0.252 | 0.252 | 0.252 | 0.252 | 0.252 | 0.252 | 0.252 | 0.252 |
| E_CCR_COAB | 15.11 | 15.11 | 15.11 | 15.11 | 15.11 | 15.11 | 15.11 | 15.11 |
| E_FGD_COABH_N | 3.184 | 3.184 | 3.184 | 3.184 | 3.184 | 3.184 | 3.184 | 3.184 |
| E_FGD_COABM_N | 2.347 | 2.347 | 2.347 | 2.347 | 2.347 | 2.347 | 2.347 | 2.347 |
| E_FGD_COABL_N | 3.797 | 3.797 | 3.797 | 3.797 | 3.797 | 3.797 | 3.797 | 3.797 |
| E_CCR_COALIGCC_N | 14.52 | 14.52 | 14.52 | 14.52 | 14.52 | 14.52 | 14.52 | 14.52 |
| E_CCR_COALSTM_N | 20 | 20 | 20 | 20 | 20 | 20 | 20 | 20 |

Table S16. List of fixed costs for all technologies except future wind and solar in M$/GW-yr, or $/kW-yr. Data sources used for cost projections are discussed in Section 3.

| Technology | Vintage | Period | | | | | | | |
|---|---|---|---|---|---|---|---|---|---|
| | | 2015 | 2020 | 2025 | 2030 | 2035 | 2040 | 2045 | 2050 |
| EBIOIGCC | ALL | 112 | 112 | 112 | 112 | 112 | 112 | 112 | 112 |
| EBIOSTMR | ALL | 12.5 | 12.5 | 12.5 | 12.5 | 12.5 | 12.5 | 12.5 | 12.5 |
| ECOALIGCC | ALL | 54.5 | 54.5 | 54.5 | 54.5 | 54.5 | 54.5 | 54.5 | 54.5 |
| ECOALIGCCS | ALL | 79.4 | 79.4 | 79.4 | 79.4 | 79.4 | 79.4 | 79.4 | 79.4 |
| ECOALSTM | ALL | 33 | 33 | 33 | 33 | 33 | 33 | 33 | 33 |
| ECOASTMR | ALL | 33 | 33 | 33 | 33 | 33 | 33 | 33 | 33 |
| EDSLCTR | ALL | 5.8 | 5.8 | 5.8 | 5.8 | 5.8 | 5.8 | 5.8 | 5.8 |
| EGEOBCFS | ALL | 119.7 | 119.7 | 119.7 | 119.7 | 119.7 | 119.7 | 119.7 | 119.7 |
| EHYDCONR | ALL | 9.7 | 9.7 | 9.7 | 9.7 | 9.7 | 9.7 | 9.7 | 9.7 |
| EHYDREVR | ALL | 14.4 | 14.4 | 14.4 | 14.4 | 14.4 | 14.4 | 14.4 | 14.4 |
| ELFGGTR | ALL | 159.2 | 159.2 | 159.2 | 159.2 | 159.2 | 159.2 | 159.2 | 159.2 |
| ELFGICER | ALL | 197.5 | 197.5 | 197.5 | 197.5 | 197.5 | 197.5 | 197.5 | 197.5 |
| ELFGICER | ALL | 197.5 | 197.5 | 197.5 | 197.5 | 197.5 | 197.5 | 197.5 | 197.5 |
| ENGAACC | ALL | 16.3 | 16.3 | 16.3 | 16.3 | 16.3 | 16.3 | 16.3 | 16.3 |
| ENGAACT | ALL | 7.5 | 7.5 | 7.5 | 7.5 | 7.5 | 7.5 | 7.5 | 7.5 |
| ENGACC05 | ALL | 14 | 14 | 14 | 14 | 14 | 14 | 14 | 14 |
| ENGACCCCS | ALL | 34.7 | 34.7 | 34.7 | 34.7 | 34.7 | 34.7 | 34.7 | 34.7 |
| ENGACCR | ALL | 4.6 | 4.6 | 4.6 | 4.6 | 4.6 | 4.6 | 4.6 | 4.6 |
| ENGACT05 | ALL | 7.8 | 7.8 | 7.8 | 7.8 | 7.8 | 7.8 | 7.8 | 7.8 |
| ENGACTR | ALL | 5.8 | 5.8 | 5.8 | 5.8 | 5.8 | 5.8 | 5.8 | 5.8 |
| ESOLPVR | ALL | 20 | 20 | 20 | 20 | 20 | 20 | 20 | 20 |
| ESOLSTCEN | ALL | 63 | 63 | 63 | 63 | 63 | 63 | 63 | 63 |
| EURNALWR | ALL | 83.4 | 83.4 | 83.4 | 83.4 | 83.4 | 83.4 | 83.4 | 83.4 |
| EURNALWR15 | ALL | 98.9 | 98.9 | 98.9 | 98.9 | 98.9 | 98.9 | 98.9 | 98.9 |
| EURNSMR | ALL | 118.7 | 118.7 | 118.7 | 118.7 | 118.7 | 118.7 | 118.7 | 118.7 |
| ESLION | ALL | 9.18 | 8.65 | 8.11 | 7.58 | 7.05 | 6.52 | 6.52 | 6.52 |
| ESZINC | ALL | 32.0 | 29.0 | 28.0 | 26.0 | 24.0 | 23.0 | 22.0 | 22.0 |
| ESCAIR | ALL | 32.0 | 30.0 | 29.0 | 27.0 | 26.0 | 26.0 | 25.0 | 25.0 |
| ESFLOW | ALL | 24.0 | 17.0 | 11.0 | 10.0 | 9.0 | 9.0 | 9.0 | 8.0 |
| E_CCR_COAB | ALL | 0.264 | 0.264 | 0.264 | 0.264 | 0.264 | 0.264 | 0.264 | 0.264 |
| E_CCR_COALIGCC_N | ALL | 0.435 | 0.435 | 0.435 | 0.435 | 0.435 | 0.435 | 0.435 | 0.435 |
| E_CCR_COALSTM_N | ALL | 0.346 | 0.346 | 0.346 | 0.346 | 0.346 | 0.346 | 0.346 | 0.346 |
| E_LNBSCR_COAB | ALL | 0.009 | 0.009 | 0.009 | 0.009 | 0.009 | 0.009 | 0.009 | 0.009 |
| E_LNBSNCR_COAB | ALL | 0.008 | 0.008 | 0.008 | 0.008 | 0.008 | 0.008 | 0.008 | 0.008 |
| E_SCR_COAB_N | ALL | 0.009 | 0.009 | 0.009 | 0.009 | 0.009 | 0.009 | 0.009 | 0.009 |
| E_SNCR_COAB | ALL | 0.008 | 0.008 | 0.008 | 0.008 | 0.008 | 0.008 | 0.008 | 0.008 |

Table S17. List of fixed costs for future onshore wind and solar PV in M$/GW-yr, or $/kW-yr. Data sources used for cost projections are discussed in Section 3.

| Technology | Vintage | | | | | | | |
|---|---|---|---|---|---|---|---|---|
| | 2015 | 2020 | 2025 | 2030 | 2035 | 2040 | 2045 | 2050 |
| ESOLPVCEN | 12 | 11 | 10 | 10 | 10 | 10 | 10 | 10 |
| ESOLPVDIS | 24.0 | 17.0 | 10.0 | 10.0 | 10.0 | 10.0 | 10.0 | 10.0 |
| EWNDON | 51.0 | 49.0 | 47.0 | 46.0 | 44.0 | 42.0 | 40.0 | 38.0 |

Table S18. List of variable costs for all technologies in this study, M$/PJ, or $/GJ.

| Technology | Vintage | 2015 | 2020 | 2025 | 2030 | 2035 | 2040 | 2045 | 2050 |
|---|---|---|---|---|---|---|---|---|---|
| EBIOIGCC | ALL | 1.549 | 1.549 | 1.549 | 1.549 | 1.549 | 1.549 | 1.549 | 1.549 |
| EBIOSTMR | ALL | 5.909 | 5.909 | 5.909 | 5.909 | 5.909 | 5.909 | 5.909 | 5.909 |
| ECOALIGCC | ALL | 2.126 | 2.126 | 2.126 | 2.126 | 2.126 | 2.126 | 2.126 | 2.126 |
| ECOALIGCCS | ALL | 2.559 | 2.559 | 2.559 | 2.559 | 2.559 | 2.559 | 2.559 | 2.559 |
| ECOALSTM | ALL | 1.316 | 1.316 | 1.316 | 1.316 | 1.316 | 1.316 | 1.316 | 1.316 |
| ECOASTMR | ALL | 1.316 | 1.316 | 1.316 | 1.316 | 1.316 | 1.316 | 1.316 | 1.316 |
| EDSLCTR | ALL | 10.233 | 10.233 | 10.233 | 10.233 | 10.233 | 10.233 | 10.233 | 10.233 |
| EGEOBCFS | ALL | 0 | 0 | 0 | 0 | 0 | 0 | 0 | 0 |
| EHYDCONR | ALL | 5.244 | 5.244 | 5.244 | 5.244 | 5.244 | 5.244 | 5.244 | 5.244 |
| EHYDREVR | ALL | 6.124 | 6.124 | 6.124 | 6.124 | 6.124 | 6.124 | 6.124 | 6.124 |
| ELFGGTR | ALL | 0 | 0 | 0 | 0 | 0 | 0 | 0 | 0 |
| ELFGICER | ALL | 0 | 0 | 0 | 0 | 0 | 0 | 0 | 0 |
| ENGAACC | ALL | 0.963 | 0.963 | 0.963 | 0.963 | 0.963 | 0.963 | 0.963 | 0.963 |
| ENGAACT | ALL | 3.053 | 3.053 | 3.053 | 3.053 | 3.053 | 3.053 | 3.053 | 3.053 |
| ENGACC05 | ALL | 1.06 | 1.06 | 1.06 | 1.06 | 1.06 | 1.06 | 1.06 | 1.06 |
| ENGACCCCS | ALL | 2.053 | 2.053 | 2.053 | 2.053 | 2.053 | 2.053 | 2.053 | 2.053 |
| ENGACCR | ALL | 1.426 | 1.426 | 1.426 | 1.426 | 1.426 | 1.426 | 1.426 | 1.426 |
| ENGACT05 | ALL | 4.549 | 4.549 | 4.549 | 4.549 | 4.549 | 4.549 | 4.549 | 4.549 |
| ENGACTR | ALL | 9.314 | 9.314 | 9.314 | 9.314 | 9.314 | 9.314 | 9.314 | 9.314 |
| ESOLPVCEN | ALL | 0 | 0 | 0 | 0 | 0 | 0 | 0 | 0 |
| ESOLPVDIS | ALL | 0 | 0 | 0 | 0 | 0 | 0 | 0 | 0 |
| ESOLPVR | ALL | 0 | 0 | 0 | 0 | 0 | 0 | 0 | 0 |
| ESOLSTCEN | ALL | 0 | 0 | 0 | 0 | 0 | 0 | 0 | 0 |
| EURNALWR | ALL | 0.459 | 0.459 | 0.459 | 0.459 | 0.459 | 0.459 | 0.459 | 0.459 |
| EURNALWR15 | ALL | 0.63 | 0.63 | 0.63 | 0.63 | 0.63 | 0.63 | 0.63 | 0.63 |
| EURNSMR | ALL | 0.756 | 0.756 | 0.756 | 0.756 | 0.756 | 0.756 | 0.756 | 0.756 |
| EWNDOFS | ALL | 0 | 0 | 0 | 0 | 0 | 0 | 0 | 0 |
| EWNDON | ALL | 0 | 0 | 0 | 0 | 0 | 0 | 0 | 0 |
| ESLION | ALL | 0.041 | 0.041 | 0.041 | 0.041 | 0.041 | 0.041 | 0.041 | 0.041 |
| EE | ALL | 1.62 | 1.42 | 1.77 | 2.16 | 2.59 | 3.13 | 3.6 | 4.15 |
| EDISTR | ALL | 6.407 | 7.018 | 7.188 | 7.348 | 7.585 | 7.781 | 8 | 8.1 |
| ETRANS | ALL | 1.935 | 2.389 | 2.547 | 2.735 | 2.953 | 3.121 | 3.235 | 3.28 |

Table S19. List of efficiencies for all technologies in this study.

| Technology | Vintage | Efficiency | Unit |
|---|---|---|---|
| EBIOIGCC | 2015-2050 | 0.253 | Dimensionless |
| EBIOSTMR | 1975-2010 | 0.219 | Dimensionless |
| ECOALIGCC | 2015 | 0.392 | Dimensionless |
| ECOALIGCC | 2020 | 0.423 | Dimensionless |
| ECOALIGCC | 2025-2050 | 0.458 | Dimensionless |
| ECOALIGCCS | 2015-2050 | 0.411 | Dimensionless |
| ECOALIGCCS_b | 2015-2050 | 0.411 | Dimensionless |
| ECOALIGCC_b | 2015 | 0.392 | Dimensionless |
| ECOALIGCC_b | 2020 | 0.423 | Dimensionless |
| ECOALIGCC_b | 2025-2050 | 0.458 | Dimensionless |
| ECOALSTM | 2015-2050 | 0.388 | Dimensionless |
| ECOALSTM_b | 2015-2050 | 0.388 | Dimensionless |
| ECOASTMR | 1960-2010 | 0.352 | Dimensionless |
| ECOASTMR_b | 1960-2010 | 0.352 | Dimensionless |
| EDSLCTR | 1975-2010 | 0.224 | Dimensionless |
| ELFGGTR | 1995-2010 | 0.3 | Dimensionless |
| ELFGICER | 2000-2010 | 0.36 | Dimensionless |
| ENGAACC | 2015 | 0.531 | Dimensionless |
| ENGAACC | 2020 | 0.535 | Dimensionless |
| ENGAACC | 2025-2050 | 0.539 | Dimensionless |
| ENGAACT | 2015 | 0.35 | Dimensionless |
| ENGAACT | 2020 | 0.373 | Dimensionless |
| ENGAACT | 2025-2050 | 0.399 | Dimensionless |
| ENGACC05 | 2015 | 0.484 | Dimensionless |
| ENGACC05 | 2020 | 0.493 | Dimensionless |
| ENGACC05 | 2025-2050 | 0.502 | Dimensionless |
| ENGACCCCS | 2015-2050 | 0.455 | Dimensionless |
| ENGACCR | 1990-2010 | 0.474 | Dimensionless |
| ENGACT05 | 2015 | 0.313 | Dimensionless |
| ENGACT05 | 2020 | 0.345 | Dimensionless |
| ENGACT05 | 2025-2050 | 0.385 | Dimensionless |
| ENGACTR | 1990-2010 | 0.248 | Dimensionless |
| EURNALWR | 1975-1985 | 1.268 [a, b] | PJe/tonneIHM |
| EURNALWR15 | 2015-2050 | 1.268 | PJe/tonneIHM |
| EURNSMR | 2015-2050 | 1.693 [c] | PJe/tonneIHM |
| E_CCR_COAB | 2015-2050 | 0.65 | Dimensionless |
| E_CCR_COALIGCC_N | 2015-2050 | 0.8 | Dimensionless |
| E_CCR_COALSTM_N | 2015-2050 | 0.7 | Dimensionless |

a.  This efficiency represents the product of burn-up rate × thermal efficiency. Burnup is the amount of thermal heat production per tonne of 'IHM', which is initial heavy metal and refers to the enriched uranium used in the reactor core.

b.  Burnup rate: 45 GWd/tonneIHM, thermal efficiency: 32.6%.

c.  Burnup rate: 70 GWd/tonneIHM, thermal efficiency: 28%.

and Solar PV Generation. *IEEE Trans. Sustain. Energy* **2016**, *7* (1), 129–138.